\documentclass[fleqn,usenatbib]{mnras}

\usepackage{newtxtext,newtxmath}

\usepackage[T1]{fontenc}
\usepackage{natbib}
\usepackage{url}
\usepackage{setspace}
\usepackage{tabularx}
\usepackage{mathtools}
\usepackage{multicol}
\usepackage{blindtext}
\usepackage{lscape}
\usepackage{rotating}
\usepackage{caption}
\usepackage{verbatim}
\usepackage{multirow}
\usepackage{color}
\usepackage{threeparttable}
\usepackage{footnote}
\bibpunct[ ]{(}{)}{,}{a}{}{,}     
\DeclareRobustCommand{\VAN}[3]{#2}
\let\VANthebibliography\thebibliography
\def\thebibliography{\DeclareRobustCommand{\VAN}[3]{##3}\VANthebibliography}

\usepackage{subfig,graphicx}	
\usepackage{amsmath}	
\usepackage{amssymb}	

\def   \aj {{\rm {AJ}}}

\def   \apj {{\rm {ApJ}}}

\def   \apjs {{\rm {ApJS}}}
\def   \apjl {{\rm {ApJL}}}

\def   \aap {{\rm {A\&A}}}

\def   \mnras {{\rm {MNRAS}}}

\def   \pasp {{\ rm {The Publications of the Astronomical Society of the Pacific}} }
\def \pasj {{\ rm {The Publications of the Astronomical Society of Japan}} }
\def \nat  {{\ rm {Nature}} }

\def 	 \kms {{\rm \,km\,s$^{-1}$}}

\def     \arcsec {{\rm $^{\prime\prime}$}}
\def     \irdcA {{\rm G018.82$-$00.28}}
\def     \irdcB {{\rm G019.27+00.07}}
\def     \irdcD {{\rm G028.53$-$00.25}}
\def     \irdcE {{\rm G028.67+00.13}}
\def     \irdcI {{\rm G038.95$-$00.47}}
\def     \irdcJ {{\rm G053.11+00.05}}
\def     \irdcC {{\rm G028.37+00.07}}
\def     \irdcF {{\rm G034.43+00.24}}
\def     \irdcG {{\rm G034.77$-$00.55}}
\def     \irdcH {{\rm G035.39$-$00.33}}

\title[SiO Emission in IRDCs]{SiO emission as a probe of Cloud-Cloud Collisions in Infrared Dark Clouds\thanks{Based on observations carried out with the IRAM 30m Telescope under projects 041-18. IRAM is supported by INSU/CNRS (France),
    MPG (Germany) and IGN (Spain).}}

\author[G. Cosentino et al.]
{G. Cosentino$^{1}$\thanks{E-mail:giuliana.cosentino@chalmers.se},
I. Jim\'{e}nez-Serra$^{2}$, J. D. Henshaw$^{3}$,  P. Caselli$^{4}$, S. Viti$^{5}$,  A. T. Barnes$^{6}$, \newauthor J.C. Tan$^{1,7}$, F. Fontani$^{8}$, B. Wu$^{9}$ \\
$^{1}$Space, Earth and Environment Department, Chalmers University of Technology, SE-412 96 Gothenburg, Sweden\\
$^{2}$Centro de Astrobiolog\'{i}a (CSIC/INTA), Ctra. de Torrej\'on a Ajalvir km 4, Madrid, Spain\\
$^{3}$Max Planck Institute for Astronomy, K\"onigstuhl 17, D-69117 Heidelberg, Germany\\
$^{4}$Max Planck Institute for Extraterrestrial Physics, Giessenbachstrasse 1, 85748 Garching bei M\"unchen, Germany\\
$^{5}$Department of Physics and Astronomy, University College London, Gower Street, London WC1E6BT, UK \\
$^{6}$Argelander-Institut f\"ur Astronomie, Universit\"at Bonn, Auf dem H\"ugel 71, 53121, Bonn, Germany\\
$^{7}$Department of Astronomy, University of Virginia, 530 McCormick Road
Charlottesville, 22904-4325 USA\\
$^{8}$INAF  Osservatorio Astronomico di Arcetri, Largo E. Fermi 5, 50125 Florence, Italy\\
$^{9}$National Astronomical Observatory of Japan, Yubinbango 181-8588 Tokio, Mitaka, Osawa 2-21-1, Japan\\}

\date{Accepted XXX. Received YYY; in original form ZZZ}

\pubyear{2020}

\begin{document}
\label{firstpage}
\pagerange{\pageref{firstpage}--\pageref{lastpage}}
\maketitle

\begin{abstract}\label{abstract}Infrared Dark Clouds (IRDCs) are very dense and highly extincted regions that host the initial conditions of star and stellar cluster formation. It is crucial to study the kinematics and molecular content of IRDCs to test their formation mechanism and ultimately characterise these initial conditions. We have obtained high-sensitivity Silicon Monoxide, SiO(2-1), emission maps toward the six IRDCs, \irdcA, \irdcB, \irdcD, \irdcE, \irdcI \space and \irdcJ \space (cloud A, B, D, E, I and J, respectively), using the 30-m antenna at the Instituto de Radioastronom\'{i}a Millim\'{e}trica (IRAM30m). We have investigated the SiO spatial distribution and kinematic structure across the six clouds to look for signatures of cloud-cloud collision events that may have formed the IRDCs and triggered star formation within them. Toward clouds A, B, D, I and J we detect spatially compact SiO emission with broad line profiles which are spatially coincident with massive cores. Toward the IRDCs A and I, we report an additional SiO component that shows narrow line profiles and that is widespread across quiescent regions. Finally, we do not detect any significant SiO emission toward cloud E. We suggest that the broad and compact SiO emission detected toward the clouds is likely associated with ongoing star formation activity within the IRDCs. However, the additional narrow and widespread SiO emission detected toward cloud A and I may have originated from the collision between the IRDCs and flows of molecular gas pushed toward the clouds by nearby H{\small II} regions.
\end{abstract}




\section{Introduction}

\label{introduction}
Infrared Dark Clouds (IRDCs) are relatively dense \citep[n(H$_2$)$\sim$ 10$^3$-10$^4$ cm$^{-3}$;][]{tan2014} and cold \citep[T$\leq$ 25 K;][]{pillai2006} regions of the sky, first detected as dark features against the mid-Infrared (IR) galactic background \citep{perault1996,egan1998}. These massive clouds show very little level of star formation activity, present H$_2$ column densities similar to those measured in known high-mass star forming regions \citep{rathborne2006,simon2006b,peretto2010} and furthermore they can host cold cores, i.e., the earliest phase of massive star formation. 
For all these reasons, in the past decade IRDCs have been proposed as the birthplace of massive stars ($\geq$ 8 M$_{\odot}$) and stellar clusters \citep{carey2000,rathborne2006,battersby2010}. It is nowadays clear that IRDCs are the densest regions of Giant Molecular Clouds \citep[e.g., ][]{barnes2018}, harbouring star formation at a wide range of stellar mass i.e., from low- to high-mass star and stellar cluster formation \citep{foster2014,sanhueza2017,pillai2019}. However, it is not entirely clear yet the mechanism that ignites the star formation process in such clouds.

\noindent
As seen by means of simulations, both the mechanisms responsible for assembling the cloud, such as flow-driven formation scenario, gravitational collapse, cloud-cloud collisions \citep{hennebelle2008,heitsch2009,nguyen2012,tasker2009,vanloo2014} and the dynamical processes that IRDCs undergo during their lifetime \citep{klessen2016,kruijssen2019}, can efficiently initiate the star formation process within the clouds. However, it is not yet entirely clear the relative importance of magnetic field, turbulence and gravity in regulating the formation of molecular dense structures at all scales \citep{commercon2011,fontani2018,Yang2019}. 
In particular, some formation models, such as the flow-driven scenario, presents major problems when fields are considered \citep{kortgen2015,kortgen2016}.


\noindent
Hence, it is crucial to investigate the formation mechanism and dynamics of IRDCs to better understand the physical process that sets in star and stellar cluster formation in such objects. Furthermore, a deep understanding of the cloud dynamics is crucial to reproduce the measured levels of star formation efficiency in galaxies \citep{leroy2008,ceverino2009}.

\noindent 
Among the different scenarios, IRDCs have been proposed to form at the shock-compressed layer within the interface of low velocity \citep[$\sim$10 \kms;][]{wu2016,wu2017a,wu2017b,li2017} collisions between pre-existing, more massive molecular clouds and/or filaments. Collisions of such pre-existing molecular structures may be induced both by their natural motion across the galactic plane \citep{tan2000,tasker2009,vanloo2014,wu2015,henshaw2013,inoue2013,jimenezserra2014,inutsuka2015,colling2018} and by dynamical processing caused by external stellar feedback e.g., induced by the expanding shells of supernova remnants (SNRs) and/or H{\small II} regions \citep{fukui2018,fukui2019,cosentino2019}.\\
\noindent
Due to the shock associated with the cloud-cloud (or filament-filament) collisions, fossil records of such interactions are expected to be found in the radial velocities, line profiles and chemistry of the molecular emission observed toward IRDCs \citep{tasker2009,nguyen2013,wu2015,wu2016,wu2017a,wu2017b,bisbas2017}. In particular, due to the relatively low velocity of the shock interaction and its extended spatial scale, molecular shock tracers, such as Silicon Monoxide (SiO), are expected to show narrow line profiles (few km s$^{-1}$) and to be widespread at parsec-scales. These features are in contrast with those typically seen in molecular outflows in sites of on-going star formation activity, where the line profiles are broad (with linewidths of several tens of km s$^{-1}$) and concentrated around the vicinity of the protostars  \citep{martinpintado1992,jimenezserra2005,jimenezserra2011,codella1999}. \\

\noindent
The first attempt to directly detect signatures of cloud-cloud collisions as the formation mechanism of IRDCs, was reported in \citet{jimenezserra2010}. These authors studied the kinematic structure and line profiles of the shock tracer SiO toward the cloud \irdcH. SiO is is an excellent shock tracer \citep{schilke1997} because it is known to be heavily depleted in quiescent regions \citep[$\chi\leq$ 10$^{-12}$;][]{martinpintado1992,jimenezserra2005} while it is dramatically enhanced in outflows (by several orders of magnitude) by the processing of dust grains in shocks \citep[][]{martinpintado1992,jimenezserra2005,jimenezserra2008,jimenezserra2009} when dust grains are processed by shocks. Toward \irdcH, \citet{jimenezserra2010} reported the detection of a bright and broad SiO component associated with sites of ongoing star formation activity together with widespread and fainter SiO emission characterised by narrow ($\leq$2 \kms) line profiles and located toward the quiescent regions of the cloud. The authors suggested, among other possibilities, that the narrow SiO emission component may be the fossil record of a cloud-cloud collision from which the IRDC has been formed. Later studies of the kinematic structure of the cloud \citep{jimenezserra2014,henshaw2014} have suggested that this may be the result of the merging of several pre-existing molecular filaments at larger scales \citep{henshaw2013}. Such a scenario has been supported by further studies of the kinematics and chemical properties of the cloud \citep{bisbas2018,tie2018,juvela2018,sokolov2019}.\\

\noindent 
In \citet{cosentino2018}, we extended the study reported by \citet{jimenezserra2010} to the three IRDCs \irdcC, \irdcF \space and \irdcG. Among these sources, we reported the presence of very narrow and widespread SiO emission (mean linewidth 1.6 \kms) toward the IRDC \irdcG. This narrow SiO component is located in a region of lower extinction of the cloud and far away from its massive cores. In a follow-up study, we used high-angular resolution observations of the SiO emission toward  \irdcG, obtained by the \textit{Atacama Large (sub)Millimetre Array} (ALMA), to show that the shock tracer emission is the result of a large-scale shock interaction triggered by the collision  of molecular gas pushed toward the IRDC by the nearby supernova remnant W44 \citep{cosentino2019}.\\
\noindent
Studies reported in \citet{jimenezserra2010} and \citet{cosentino2018}, highlight single-dish observations of SiO as an useful tool to test the formation mechanism and large-scale dynamics of IRDCs through cloud-cloud collisions. In this paper, we aim to extend the study of the SiO emission to six additional IRDCs: \irdcA, \irdcB, \irdcD, \irdcE, \irdcI \space and \irdcJ. Specifically, we attempt to identify signatures of cloud-cloud collisions that may have formed the clouds themselves, initiating the process of star formation within them. In Section~\ref{sample} we discuss the target selection. In Section~\ref{observations} we describe the observing method and data acquisition. In Section~\ref{method} we describe procedure and assumptions adopted to perform the data analysis. In Section~\ref{results} we present the results obtained from the analysis of the SiO emission toward the sources \irdcA, \irdcB, \irdcD, \irdcE, \irdcI, \irdcJ. In Section~\ref{discussion}, we discuss the obtained results in light of cloud formation theories and compare them with previous studies. In Section~\ref{types}, we introduce the possibility of different types of cloud-cloud collisions and discuss their relative importance for cloud and massive star formation. Finally, in Section~\ref{conclusions} we summarise our conclusions.

\section{The IRDC Sample}\label{sample}
The six IRDCs studied in this work, along with the sources \irdcH, \irdcC, \irdcF \space and \irdcG, have been presented as a ten clouds sample in \cite{butler2009,butler2012}. The ten sources are part of an extended catalogue of IRDCs, identified as dark features against the diffuse mid-IR galactic background by \citet{simon2006a}. For all the clouds of the catalogue, \citet{simon2006a} estimated the V$_{LSR}$ and kinematic distances from observations of the $^{13}$CO emission. Subsequently, \citet{rathborne2006} estimated cloud masses from the dust continuum at 1.2 mm for a sub-sample of 38 IRDCs. Masses for the ten clouds have also been estimated by \citet{kainulainen2013} from their MIR and NIR 8 $\mu$m emission, as obtained by \textit{Spitzer}. The ten clouds presented in \citet{jimenezserra2010, cosentino2018} and this paper, were selected by \citet{butler2009} from the 38 cloud sample in \citet{rathborne2006} for being located relatively nearby (kinematic distance $\leq$ 6 kpc), for being relatively massive (0.2-29 $\times$10$^3$ M\sun) and/or for showing the highest levels of contrast against the diffuse Galactic background emission at 8 $\mu$m. The cloud morphology varies across the sample with clouds A, D and I being more filamentary and clouds B, E and J showing more globular shapes. 

\section{Observations}\label{observations}
The J$=$2$\rightarrow$1 rotational transition of SiO ($\nu$=86.84696 GHz) was mapped in July 2017 toward the six IRDCs \irdcA, \irdcB, \irdcD, \irdcE, \irdcI, \irdcJ \space \citep[thereafter cloud A, B, D, E, I and J;][]{butler2009} using the 30m single dish antenna at Instituto de Radioastronomia Millimetrica (IRAM 30m, Pico Veleta, Spain). Observations were performed in On-The-Fly (OTF) observing mode with angular separation in the direction perpendicular to the scanning direction of 6\arcsec. Central coordinates, off positions and map sizes adopted for the six sources are listed in Table~\ref{tab1}. For the observations, we used the FTS spectrometer set to provide a spectral resolution of 50 kHz, corresponding to a velocity resolution of 0.16 \kms \space at the SiO rest frequency. Intensities were measured in units of antenna temperature, T$^{*}_{A}$, and converted into main-beam brightness temperature, T$_{mb}$= T$^{*}_{A}$ (F$_{eff}$/B$_{eff}$), using beam and forward efficiencies of B$_{eff}$=0.81 and F$_{eff}$=0.95, respectively. The final data cubes were created using the {\sc CLASS} software within the {\sc GILDAS} package\footnote{See http://www.iram.fr/IRAMFR/GILDAS.} and have a spatial resolution of 30\arcsec \space and a pixel size of 15\arcsec$\times$15\arcsec. In order to achieve this we convolved the native resolution data with a Gaussian kernel of $\theta$ = 10\arcsec (HPBW). The rms achieved during observations is $\sim$10 mK per 0.16 \kms \space channel but all spectra were smoothed in velocity to improve the signal-to-noise ratio of the measured line emission. This provides a final velocity resolution (i.e. channel width) of $\delta$V = 0.3 \kms.

\begin{table*}
\begin{threeparttable}
    \caption{Names, central coordinates, velocities with respect to the Local standard rest (V$_{LSR}$), kinematic distances, effective radii and masses of the six IRDCs. For each cloud, the offset position, map size and achieved rms are also reported.}
    \begin{tabular}{llllllllll}
    \hline
    \hline
         Cloud & \multicolumn{2}{c}{Central Coordinates$^a$} & V$_{LSR}^b$ & d$^b$ & R$_{eff}^c$ & Mass$^d$  &Off Position & Map Size & rms  \\
               & RA(J2000)& DEC(J2000) & [\kms] & [kpc] & [pc] & [10$^3$ M$_{\odot}$] & [\arcsec,\arcsec] & [\arcsec$\times$\arcsec] & [mK] \\
    \hline
(A) \irdcA &18:26:18.7 &-12:41:16.3 &65.8 &4.8 &10.4 &18.5 &$-$200,+100 &300$\times$240 &5 \\
(B) \irdcB &18:25:56.1 &-12:04:47.2 &26.2 &2.4 &2.7	&2.2 &+130,+300 &250$\times$230 &7\\
(D) \irdcD &18:44:17.4 &-04:00:31.4 &87.0 &5.7 &16.9 &74.3 &+200,+40 &330$\times$360 &4\\
(E) \irdcE &18:43:08.1 &-03:44.54.3 &79.5 &5.1 &11.5 &28.7 &+110,+380 &180$\times$160 &9\\
(I) \irdcI &19:04:08.1 &+05:09:15.0 &41.6 &2.7 &3.7 &2.7 &$-$120,+190 &180$\times$180 &6\\
(J) \irdcJ &19:29:16.7 &+17:54:40.0 &22.0 &1.8 &0.8 &0.2 &+190,$-$10  &160$\times$170 &5\\
\hline
\end{tabular}
\begin{tablenotes}
\item $^a$\cite{butler2009}.  $^b$\cite{rathborne2006}. $^c$ \cite{butler2012}. $^d$ Mass estimated by \citet{kainulainen2013} from combined Mid and Near IR extinction maps.
\end{tablenotes}
\label{tab1}
\end{threeparttable}
\end{table*}

\section{Method and Assumptions}\label{method}
\begin{figure*}
    \centering
    \includegraphics[scale=0.4]{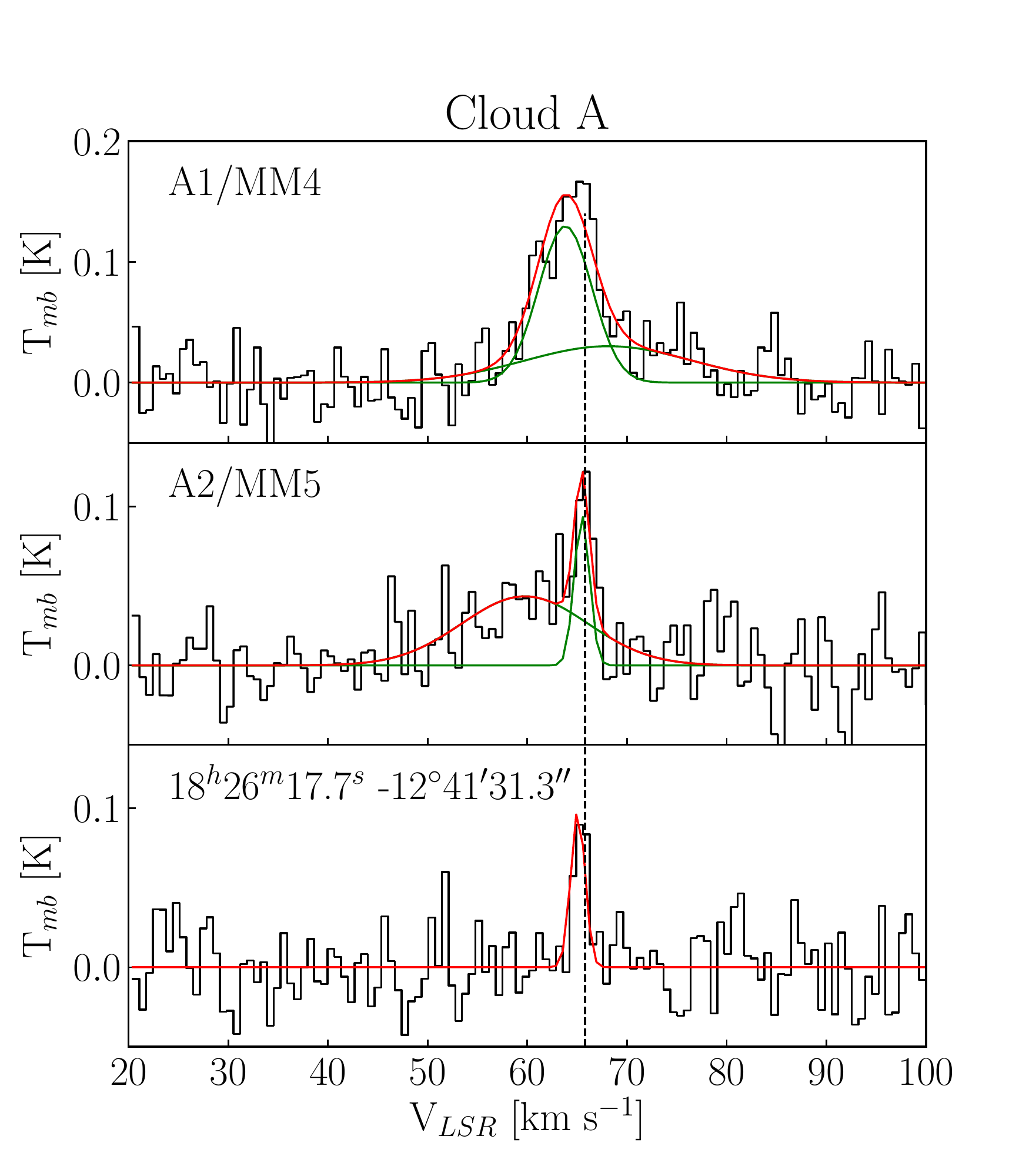}\includegraphics[scale=0.4]{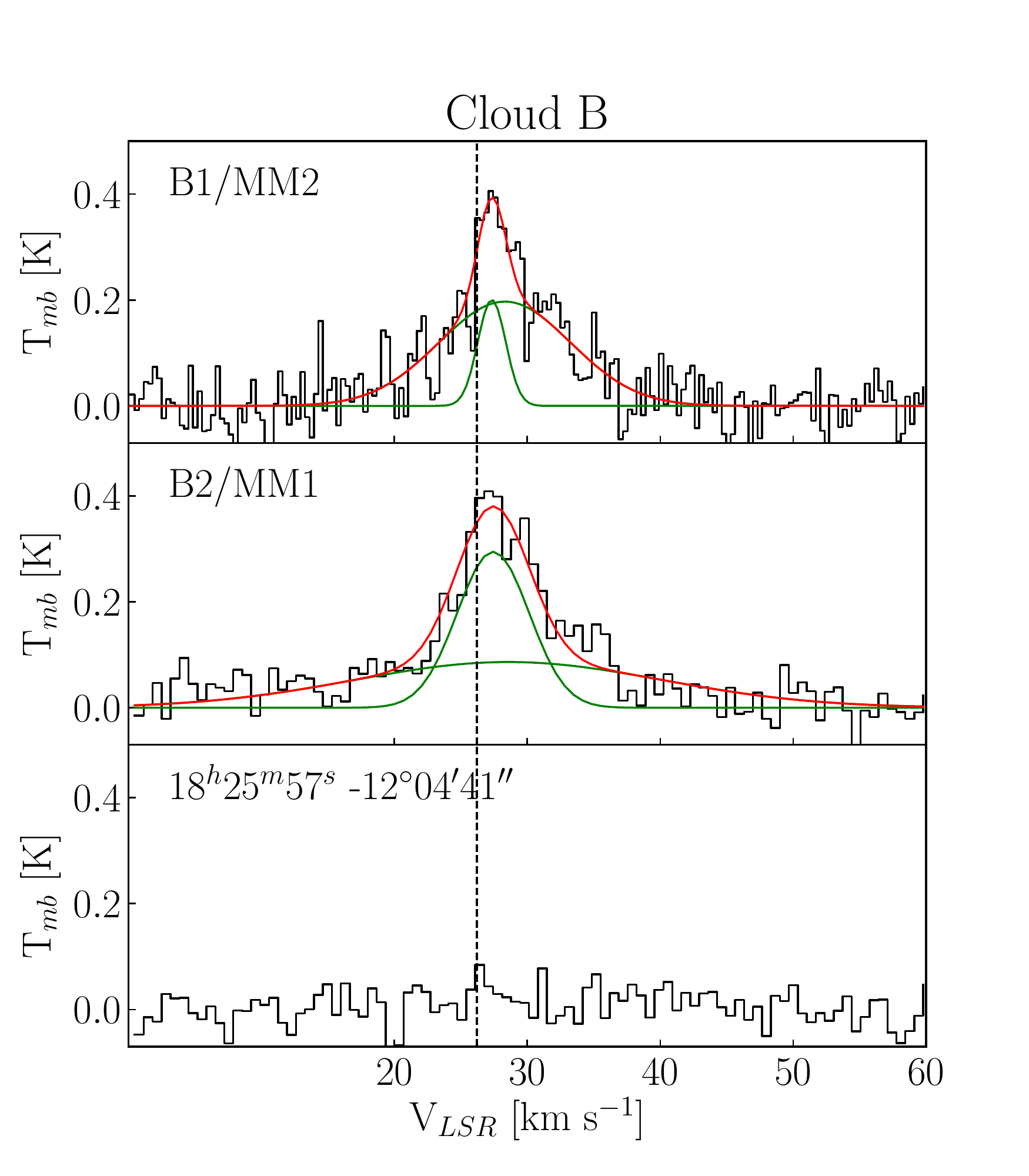}
    \includegraphics[scale=0.4]{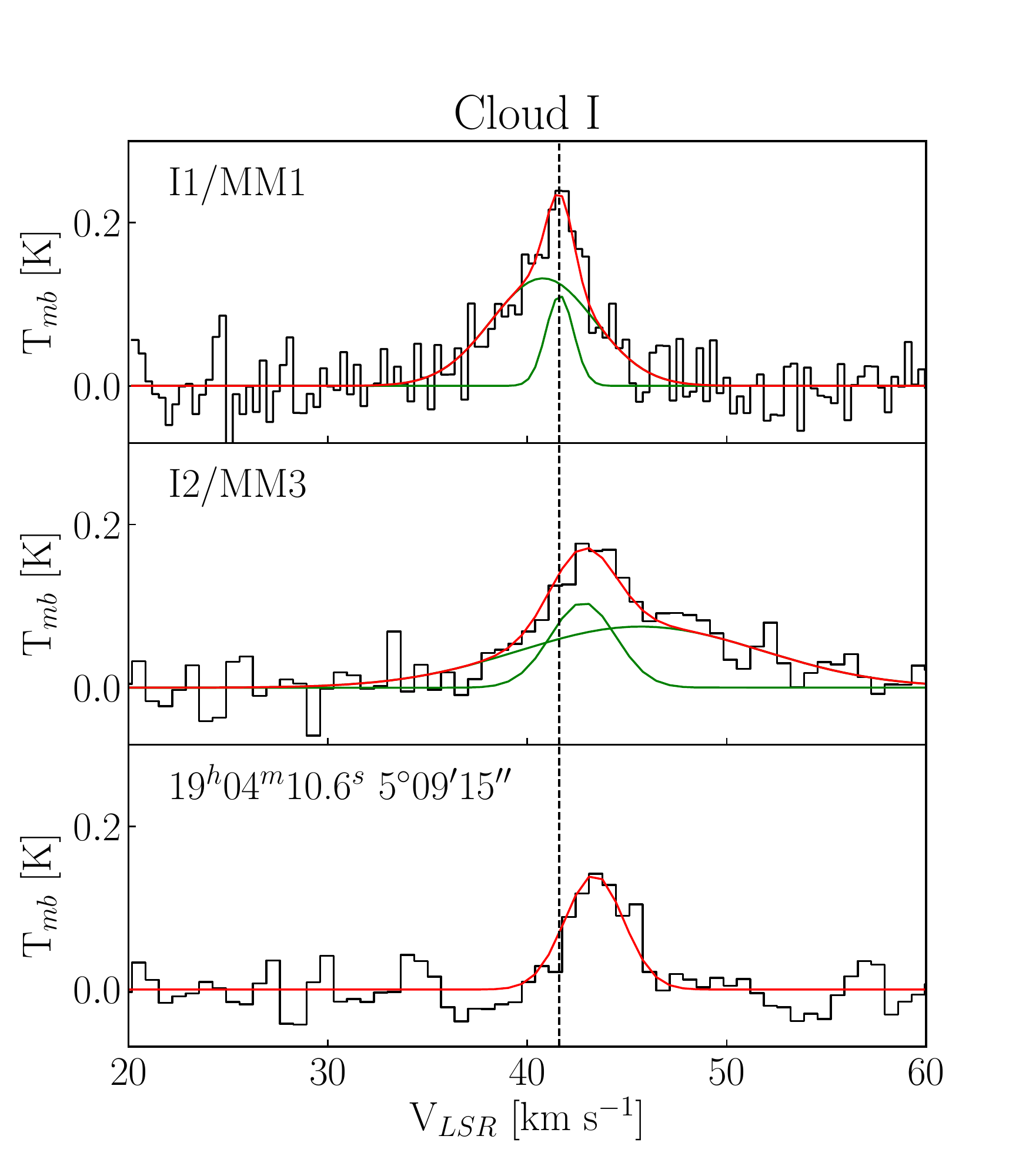}\includegraphics[scale=0.4]{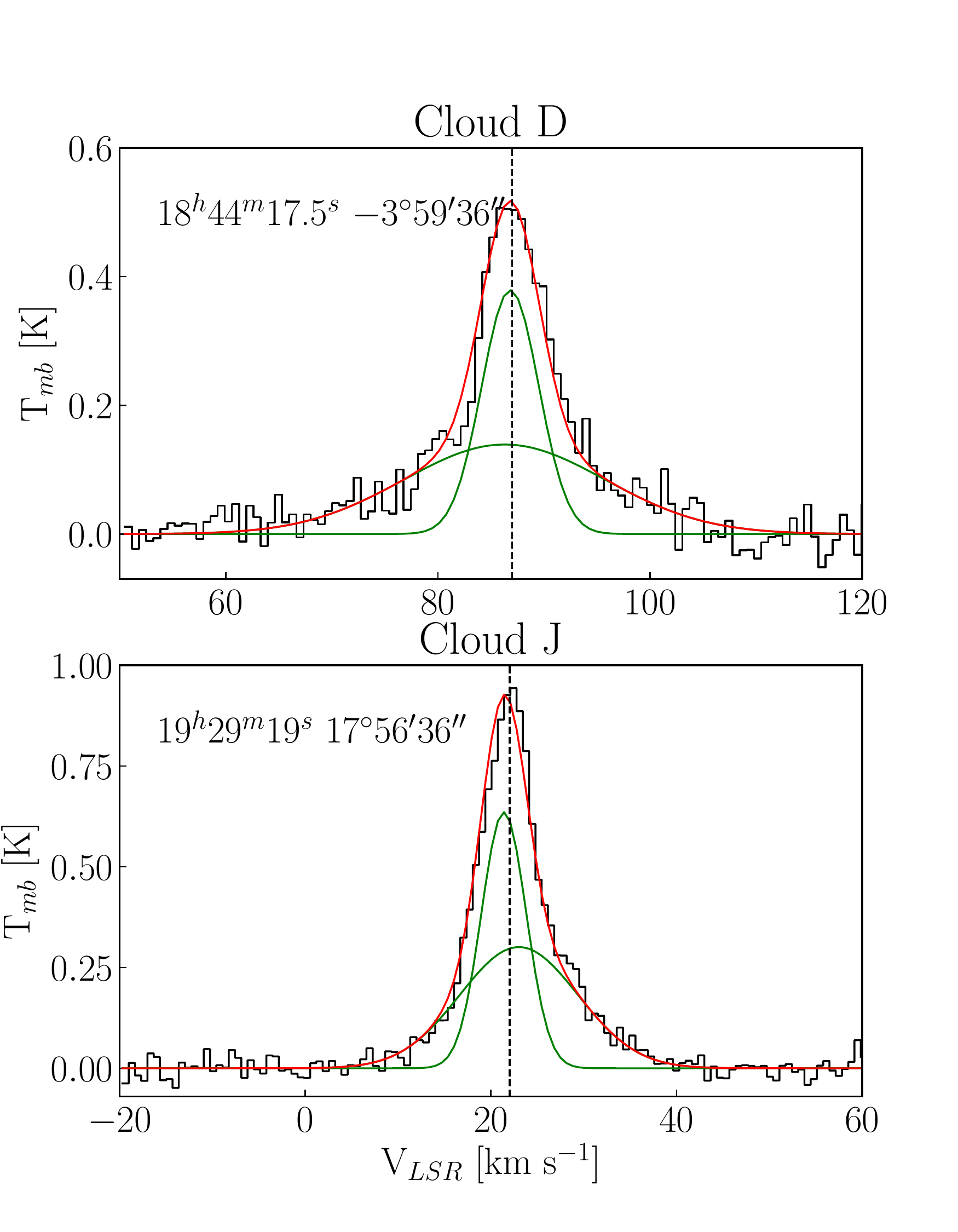}
    \caption{SiO spectra extracted toward selected positions (indicated in each panel) across cloud A (top left), B (top right), I (bottom left), D and J (bottom right) across a beam aperture of 30\arcsec. In all panels, the multi-Gaussian fitting is indicated as a red line, while the single Gaussian components are indicated as green lines. For all clouds, the corresponding central velocity is indicated as a vertical dashed line.}
    \label{SiOspectra}
\end{figure*}

In this paper, we aim to investigate the presence of differentiation in the linewidth and velocity distribution of the SiO emission across the different positions, toward the six IRDCs. In particular, the ultimate aim of this study is to identify the presence of narrow and widespread SiO emission across the six sources. This is similar to the study performed by \citet{cosentino2018} toward clouds \irdcC, \irdcF \space and \irdcG \space (corresponding to clouds C, F and G in the Butler \& Tan 2009 sample).\\
\noindent
In Figure~\ref{SiOspectra}, we show SiO spectra extracted across clouds A, B, D, I and J toward several positions of both star forming and quiescent components. Since no significant SiO emission is detected toward cloud E (see below), we do not show spectra extracted toward this source. The spectra have been extracted over a beam aperture of 30\arcsec and the corresponding positions are indicated as red diamonds in Figure~\ref{siomaps}.

\noindent
As shown in Figure~\ref{SiOspectra}, the line profile of the SiO emission across the six clouds shows a complex structure (red curves) i.e. with multiple velocity components (green curves)showing different linewidths and peak intensities. Motivated by the complex line profiles of the SiO emission across the IRDCs, we use the {\sc IDL} tool {\sc SCOUSE}\footnote{ https://github.com/jdhenshaw/SCOUSE} \citep{henshaw2016} to perform a Gaussian deconvolution of all the spectra. {\sc SCOUSE} provides a fast, robust and systematic method to perform multi-Gaussian fitting of all the spectra stored in a datacube, allowing the user to obtain information on the central velocity, peak intensity and linewidth of every single emission line above the (user-defined) detection level. In our analysis, we consider as significant all lines having intensity I$>$3$\times$rms. Moreover for each identified Gaussian component, we also require that the area underneath the curve fulfils the following condition:

\begin{equation}
    A \geq 3 \times rms \times \sqrt{\delta V \Delta V}
\label{arms}    
\end{equation}

\noindent
Where $A$ is the area of the Gaussian component and the right-hand side of the equation is the 3$\times$rms integrated over a velocity range equal to the line width of the Gaussian. $\delta V$ and $\Delta V$ correspond to the spectral velocity resolution and the line FWHM, respectively.\\ 
In addition, we have set the tolerance parameters within {\sc SCOUSE} so that linewidths computed by the code are always larger than the velocity resolution in the spectra. Moreover, we impose that the separation in centroid velocity between two adjacent Gaussian components must be $>$0.5$\Delta$V$_{min}$, where $\Delta$V$_{min}$ is the narrower linewidth of the two Gaussian components. Finally, {\sc SCOUSE} ensures the uniqueness of the results by applying post-fitting controls that are described in details in Section 3.1.5 of \citet{henshaw2016}.\\

\noindent 
From the information provided by the {\sc SCOUSE} output, we produce histograms showing the linewidth and velocity distributions of the SiO emission across the six IRDCs. Hence, from the obtained distributions, we investigate the presence of differentiation in the line profile features across a map. In particular, we use the velocity distributions to study the kinematic structures of the shocked gas across the clouds. Thus, we use the line width distributions to detect the presence of narrow SiO emission across the IRDCs. We note that {\sc SCOUSE} provided us uncertainty on the line widths and centroid velocities of the Gaussian fitting of $\sim$0.5 \kms \space and $\sim$0.3 \kms, respectively. This indicates that the scattered components in the linewidth and centroid distributions in Figure~\ref{SiOwidth} and ~\ref{SiOvelo} are not artefacts due to the goodness of the fitting.\\

\noindent
In order to disentangle between the spectrally narrow and broad SiO emission, we defined a linewidth threshold of 5 \kms. This is consistent with the method described in \citet{cosentino2018}, where all emission with linewidths below 5 \kms \space was defined as narrow. This is justified by the fact that the maximum linewidth observed for the dense gas tracer H$^{13}$CO$+$ and HN$^{13}$C emission in clouds C, F and G is 5 \kms. From a preliminary analysis (Cosentino et al. in prep), this seems to be the case also for the H$^{13}$CO$+$ and HN$^{13}$C emission toward clouds A, B, D, E, I and J. Hence, toward the ten clouds of the sample, these dense gas tracers mainly probe the dense ambient gas in our sources and there is no evidence that shocks affect their linewidth \citep[][; Cosentino et al. in prep.]{cosentino2018}. Hence, also in this work, we set a threshold of 5 \kms \space to disentangle between narrow (linewidth $\leq$ 5 \kms) and broad (linewidth $>$ 5 \kms) line profiles in the SiO emission.\\
\noindent 
Finally, in \citet{cosentino2018}, we adopted a histogram bin-size for the velocity and line width distributions corresponding to 1/3 of the mean intensity-weighted line widths of the dense gas emission. The H$^{13}$CO$+$ and HN$^{13}$C mean intensity-weighted line width is $\sim$1.5 \kms \space in cloud C, F and G. Hence we used a bin size 0.5 \kms \space for all the histograms of the three clouds. This was to allow a direct comparison between the kinematic structure of the shocked gas (SiO emission) and that of the more quiescent dense gas (H$^{13}$CO$+$ and HN$^{13}$C emission). From a preliminary investigation, the mean intensity-weighted line widths measured for the dense gas tracers toward clouds A, B, D, E, I and J  is $\sim$1.8 \kms \space (Cosentino et al. in prep.). Hence, following the method adopted in \citet{cosentino2018}, a bin size of 0.6 \kms \space should be employed here to build the SiO line width and velocity distributions. However, in this work, we will still use the slightly smaller bin size of 0.5 \kms, in order to allow a direct comparison with the results obtained for cloud C, F and G. 

\section{Results}\label{results}
\subsection{The SiO Spatial Distribution: Looking for Widespread Emission}
\begin{figure*}
    \centering
    \includegraphics[scale=0.28,trim = 0cm 2cm 1.5cm 3.5cm, clip=True]{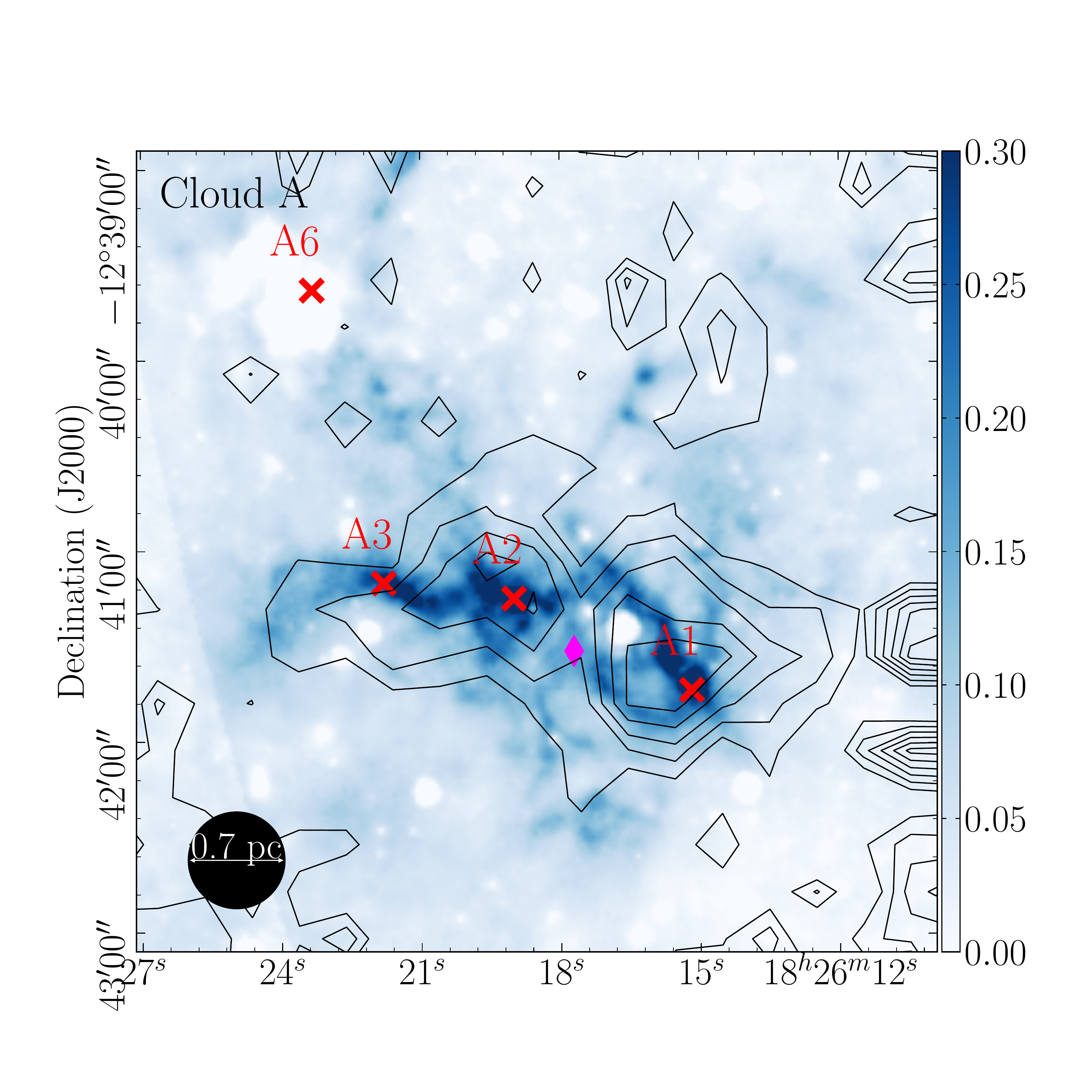}\includegraphics[scale=0.28,trim = 0cm 2cm 0.7cm 3.5cm, clip=True]{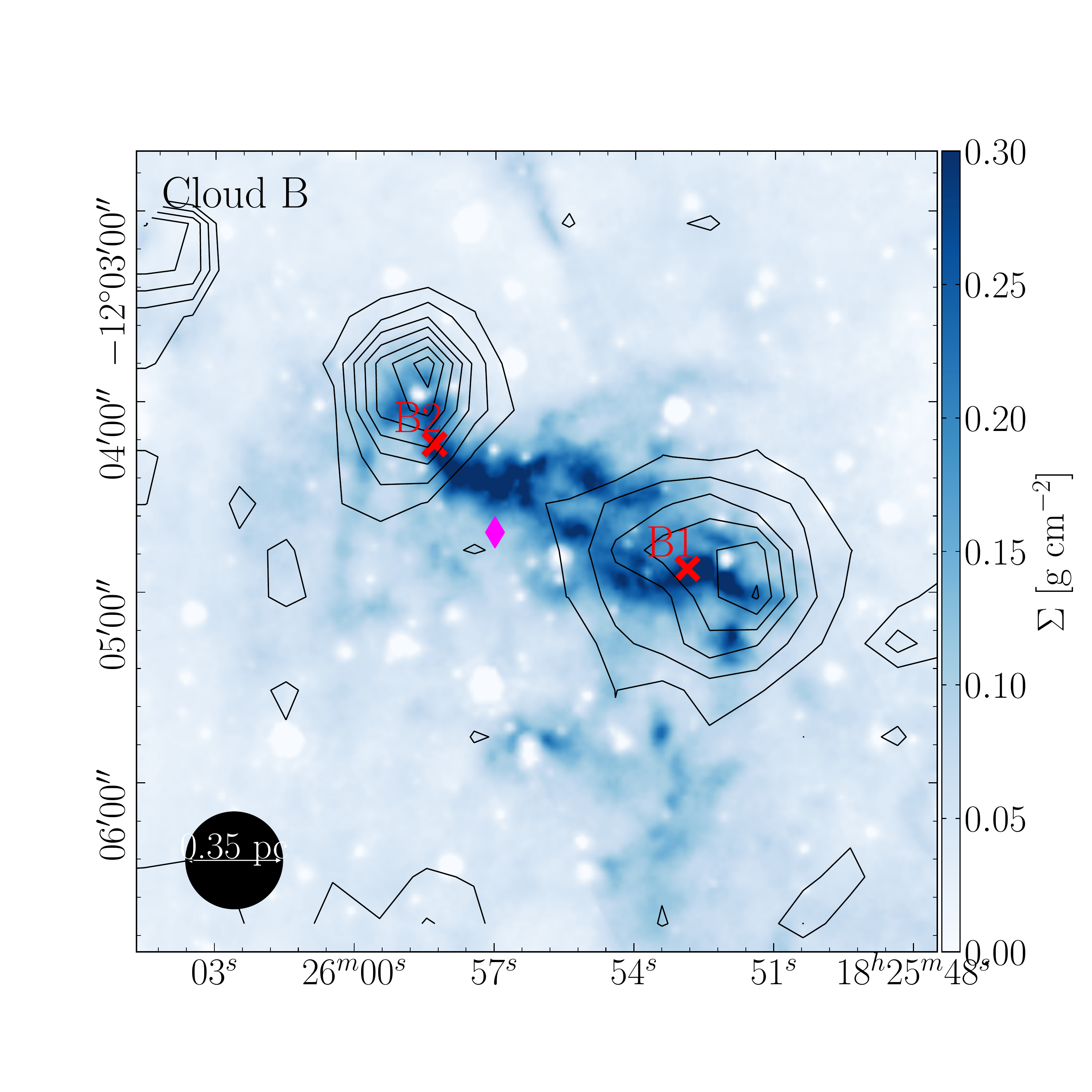}
    \includegraphics[scale=0.28,trim = 0cm 2cm 1.5cm 3.5cm, clip=True]{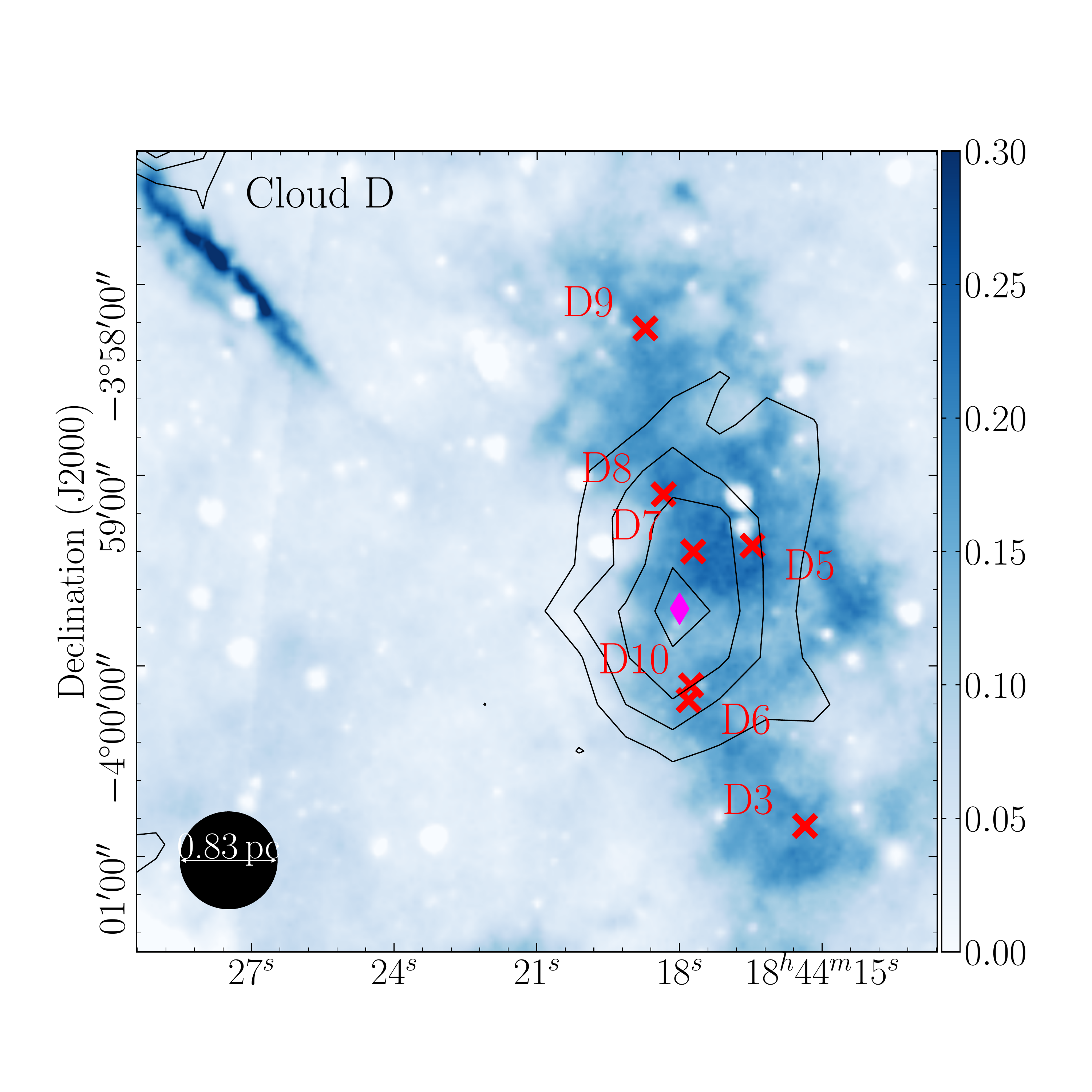}\includegraphics[scale=0.28,trim = 0cm 2cm 0.7cm 3.5cm, clip=True]{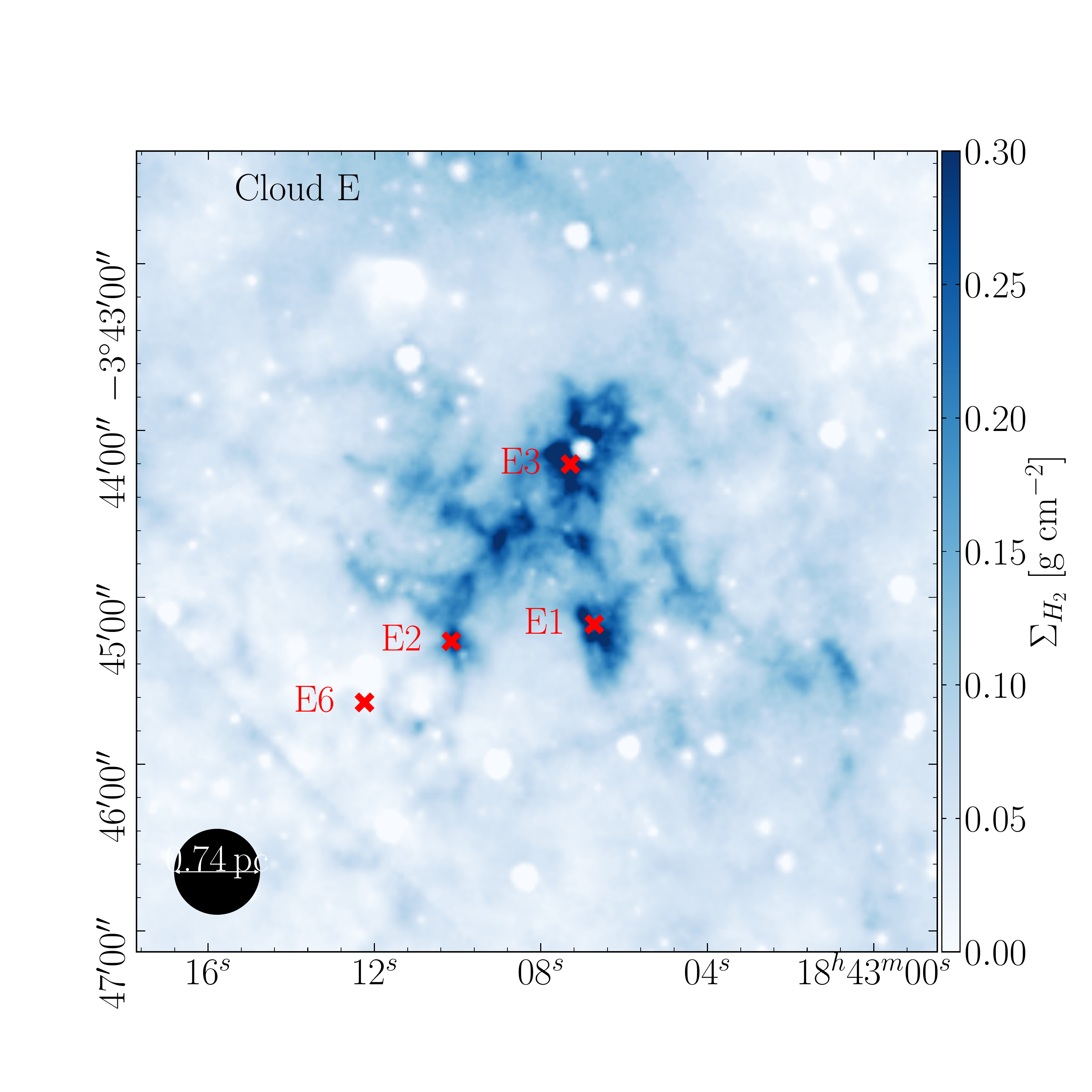}
    \includegraphics[scale=0.28,trim = 0cm 0cm 1cm 3.5cm, clip=True]{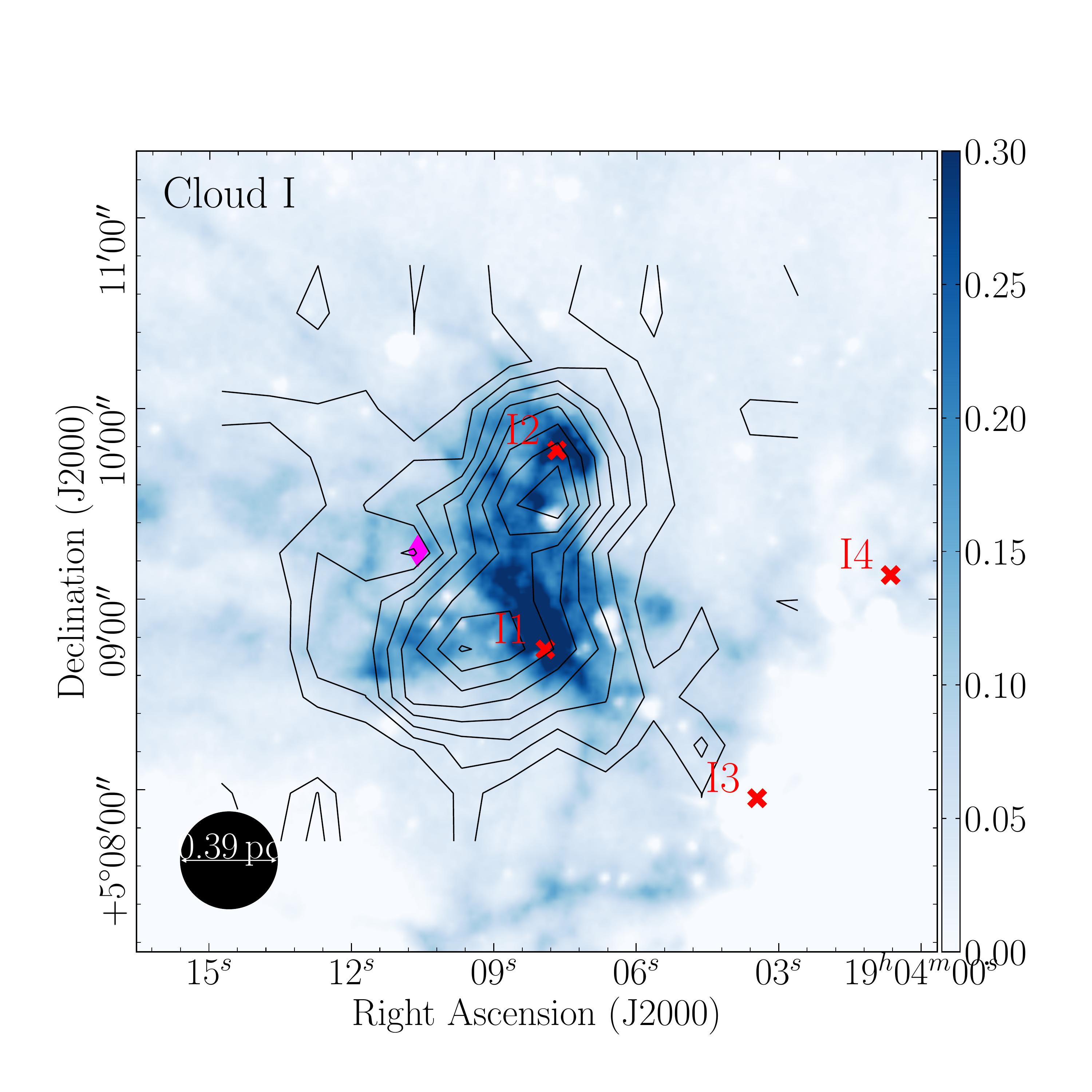}\includegraphics[scale=0.28,trim = 0cm 0cm 1cm 3.5cm, clip=True]{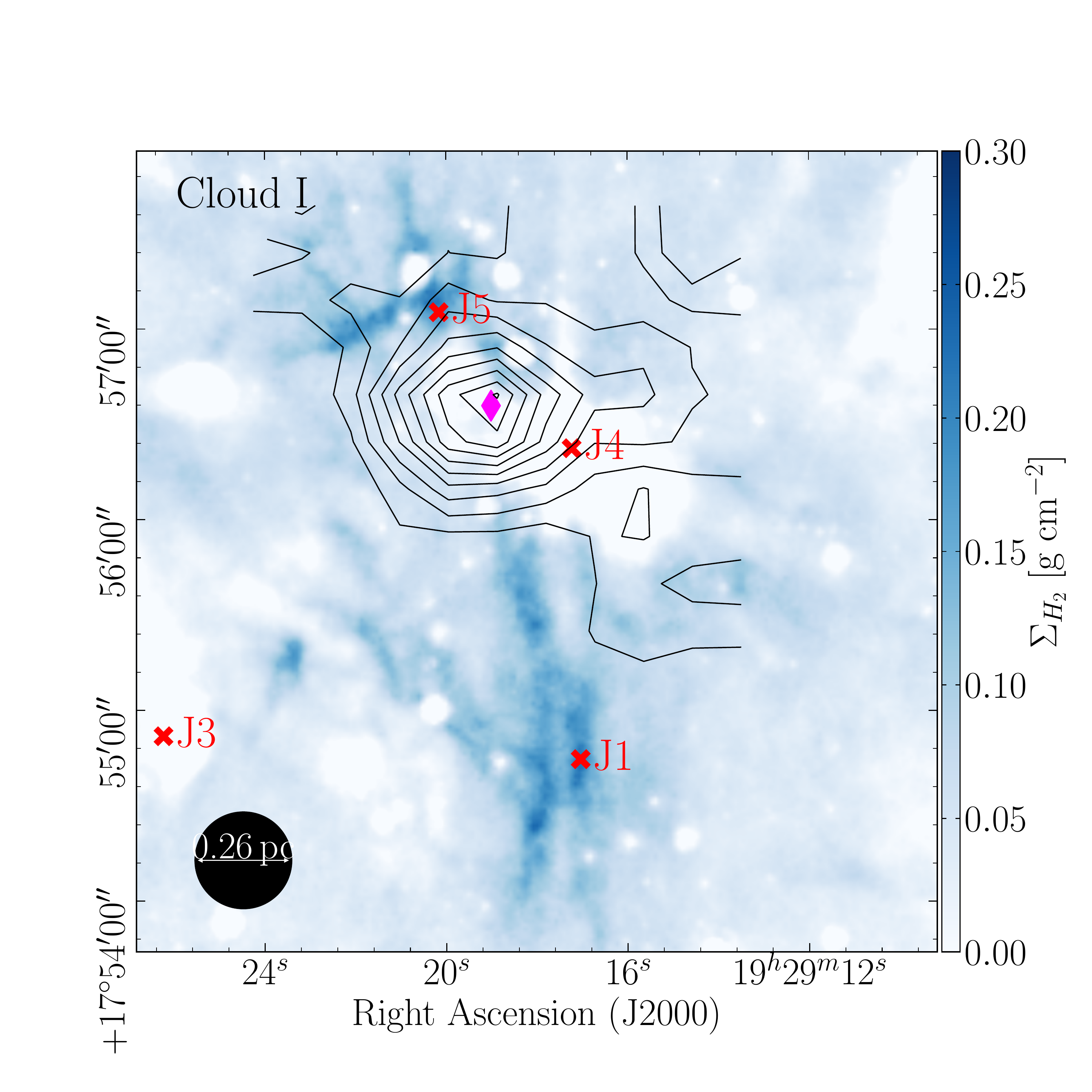}
    \caption{SiO(2-1) integrated intensity maps towards clouds A (top left panel), B (top right panel), and D (middle left panel), E (middle right panel), I (bottom left panel) and J (bottom right panel). Emission levels (black contours) are from 3$\sigma$ to 30$\sigma$ by step of 6$\sigma$ for cloud J and by steps of 3$\sigma$ for all the other IRDCs. The contours are superimposed on the mass surface density maps (grey scale) obtained by \citet{kainulainen2013}. The integration ranges are 40 to 100 \kms, 5 to 45 \kms, 4 to 120 \kms, 10 to 70 \kms \space and $-$10 to 60 \kms \space for cloud A, B, D, I and J respectively. $\sigma$ = 0.1 K \kms \space for cloud A, E and I, $\sigma$=0.2 K \kms \space for cloud B and J and $\sigma$ = 0.5 K \kms \space for cloud D. The core positions \citep[black crosses;][]{butler2009,butler2012} and the beam sizes (black circles) are shown in all panels.}
    \label{siomaps}
\end{figure*}

We investigate the spatial distributions of the SiO emission across the six clouds of the sample and report in Figure~\ref{siomaps} the SiO integrated intensity maps for cloud A, B, D, E, I and J. The emission levels (black contours) are superimposed on the mass surface density maps (in blue scale) obtained by \citet{kainulainen2013} and in these and all the following maps, the names and positions \citep[black crosses;][]{rathborne2006,butler2009,butler2012} of the massive cores within the clouds are indicated.\\
\noindent
We detect very bright and extended SiO emission toward cloud A, B, D, I and J and no emission above the 3$\sigma$ level ($\sigma$ = 0.07 K \kms) toward cloud E. The emission across cloud A (integration range 40 to 100 \kms) and I (integration range 10 to 70 \kms) shows similar features i.e., it is widespread across the whole filamentary structure with emission peaks in correspondence of the massive cores A1 and A2 toward cloud A and I1 and I2 toward cloud I. The shock tracer emission covers a spatial scale of 4.2$\times$2.2 parsecs$^2$ toward cloud A \citep[d$\sim$4.8 kpc;][]{simon2006b} and 1.3$\times$1.7 parsecs$^2$ toward cloud I \citep[d$\sim$2.7 kpc;][]{simon2006b}. Toward cloud B (integration range 5$-$45 \kms) the SiO is distributed among two blob-like structures spatially coincident with the two cores B1 and B2. The emission morphology is very compact, with the most extended structure covering a spatial scale of 0.3$\times$0.7 parsecs$^2$ \citep[d$\sim$2.4 kpc;][]{simon2006b} and no emission is detected toward quiescent regions across the cloud. Finally, toward cloud D (integration range 40$-$120 \kms) and J (integration range -10$-$60 \kms), very compact SiO emission is found at the centre of the regions crowded with massive cores. The SiO emission is extended across a spatial scale of 0.7$\times$2.5 parsecs$^2$ toward cloud D \citep[d$\sim$5.7 kpc;][]{simon2006b} and 0.3$\times$0.8 parsecs$^2$ toward cloud J \citep[d$\sim$1.8 kpc;][]{simon2006b}. 
  
\subsection{The SiO Line Profile: Looking for Narrow Shock Tracer Emission}
From the analysis performed with \textsc{SCOUSE}, we extract information on the line widths, central velocities and peak intensities of the SiO emission lines at each positions (averaging over a beam) across the cloud areas and build distributions of such quantities to study changes in the SiO line profiles across the IRDCs. For all histograms reported in the following Sections, we adopt a bin size of 0.5 \kms, as discussed in Sec.~\ref{method} and use the y-axis to show the percentage of emission lines having a certain line width or central velocity and normalised for each cloud to the total number of positions in which SiO emission has been detected. Since no emission is detected toward cloud E, we exclude the cloud from the following analysis and only show distributions obtained for the remaining five clouds. 

\subsubsection{The SiO line width distributions}
\begin{figure}
\centering
\includegraphics[scale=0.8,trim = 1cm 1.5cm 10cm 3cm, clip=True]{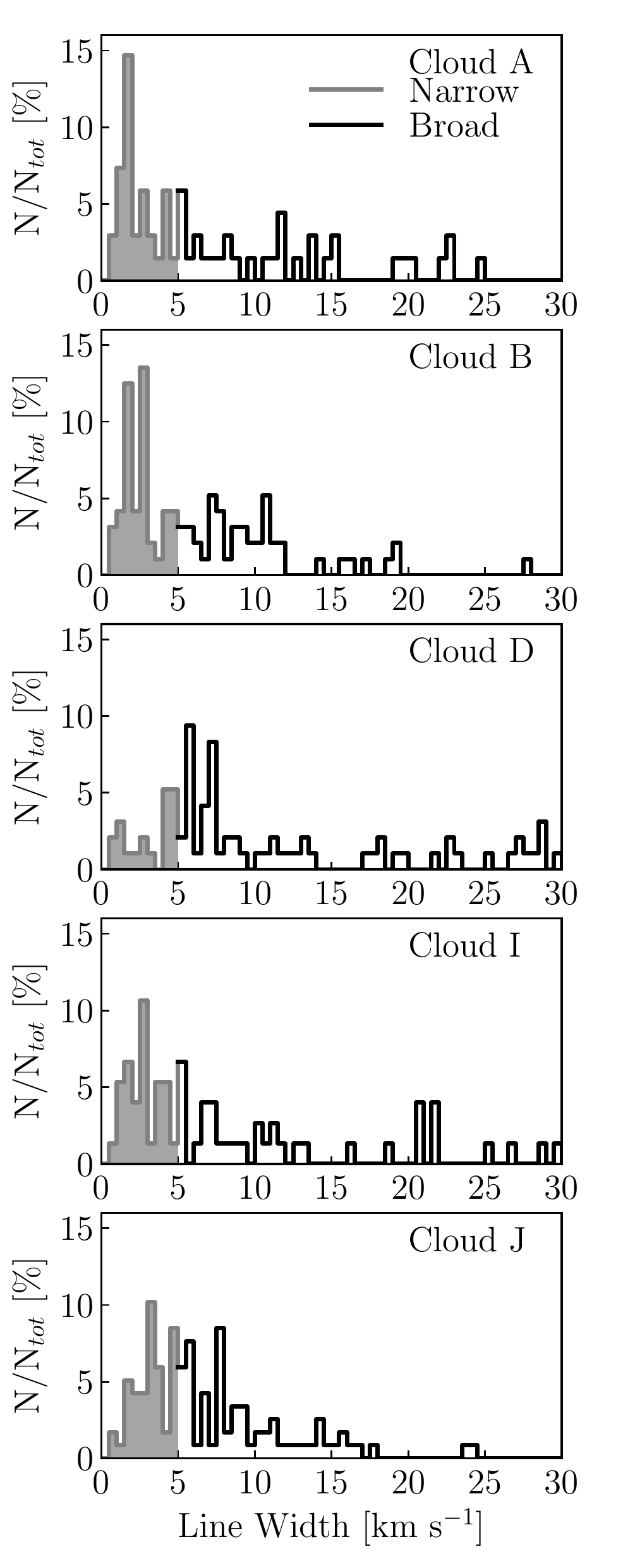}
\caption{Line width distributions of the SiO emissions obtained for cloud A, B, D, I and J. The histograms show the percentage of emission lines having line widths falling within each bin and normalised to the total number of lines detected across each cloud. Bin size is 0.5 \kms corresponding to 1/3 of the mean intensity-weighted line width obtained for the dense gas tracers in \citet{cosentino2018}. The SiO emission narrower than 5 \kms \space is indicated as grey filled histograms while the broad emission ($>$ 5\kms) is shown as empty black histograms.}
\label{SiOwidth}
\end{figure}

We now investigate variations in the SiO line widths across the five IRDCs in which shocked gas has been detected, using the SiO line width distributions obtained for clouds A, B, D, I and J (from top to bottom) as shown in Figure~\ref{SiOwidth}. As discussed in Section~\ref{method}, we adopt a threshold of 5 \kms to differentiate between broad and narrow SiO emission.\\

\noindent
Toward the five clouds, the SiO emission shows complex line width distributions with both narrow ($\Delta$V $\leq$5 \kms; grey filled histograms) and broad ($\Delta$V $>$5 \kms; black empty histograms) line width components that coexist with different relative percentages. In clouds A and B, the two components are almost equally distributed, representing each $\sim$50\% of the total SiO emission lines detected. The narrow component is clearly identified by well-defined peaks at linewidth $\sim$2 \kms, while the broad component shows a spread distribution with line widths up to $\sim$20 to 25 \kms. In cloud D, the distribution is dominated by the broad component that accounts for $80\%$ of the total detected lines, and for the narrow emission, there is no preferred peak in the narrow line width distribution. Finally, in cloud I and J, the SiO emission shows line width distributions intermediate between the case of cloud A and B and the case of cloud D. The narrow and broad emission components represent each $\sim$50\% of the total emission lines, similarly to the case of clouds A and B. However, as in cloud D, no dominant peaks are present in the narrow component distribution. We report the percentage of SiO emission having line width $\leq$3 and $\leq$5 \kms \space in Table~\ref{SiOperc}, along with the mean intensity-weighted line widths of the narrow ($\leq$5 \kms) and broad SiO emission components. For comparison, the mean intensity-weighted line width measured in \citet{cosentino2018} for cloud G is 1.6 \kms. We note that, as discussed in Section~\ref{method}, all the  line components fitted with {\sc SCOUSE} have integrated areas above 3 times the integrated noise over the same linewidth (see Eq.~\ref{arms}).

\begin{table}
    \centering
     \caption{Frequency of detection, in percentages, of the SiO narrow emission towards the six IRDCs for thresholds $\leq$3 and 5 \kms, normalised with respect to the total SiO emission. The mean intensity-weighted line widths of the narrow ($\leq$5 \kms) and broad components are also indicated, as $\langle\Delta$V$_{n}\rangle$ and $\langle\Delta$V$_{b}\rangle$ respectively.}
    \begin{tabular}{lcccc}
    \hline
    \hline
    Cloud & $\leq$3 \kms & $\leq$5 \kms & $\langle\Delta$V$_{n}\rangle$& $\langle\Delta$V$_{b}\rangle$\\
    			& [\%]	&[\%] & [\kms] & [\kms]\\
    \hline
    A & 36.8 &51.5 & 2.1 & 12.0\\
    B & 39.6 &52.1 & 2.4 &10.7\\
    D & 10.4 &22.9 & 3.1 &14.5\\
    E & $-$ & $-$  &$-$ &$-$\\
    I & 29.3 &48.0 &2.8  &12.2\\
    J & 26.3 &48.3 &3.5 &8.5\\
    \hline
    \end{tabular}
    \label{SiOperc}
\end{table}

\subsubsection{Spatial distributions of the SiO line width components}
\begin{figure*}
    \centering
    \includegraphics[scale=0.25,trim = 0.3cm 0cm 0cm 0cm, clip=True]{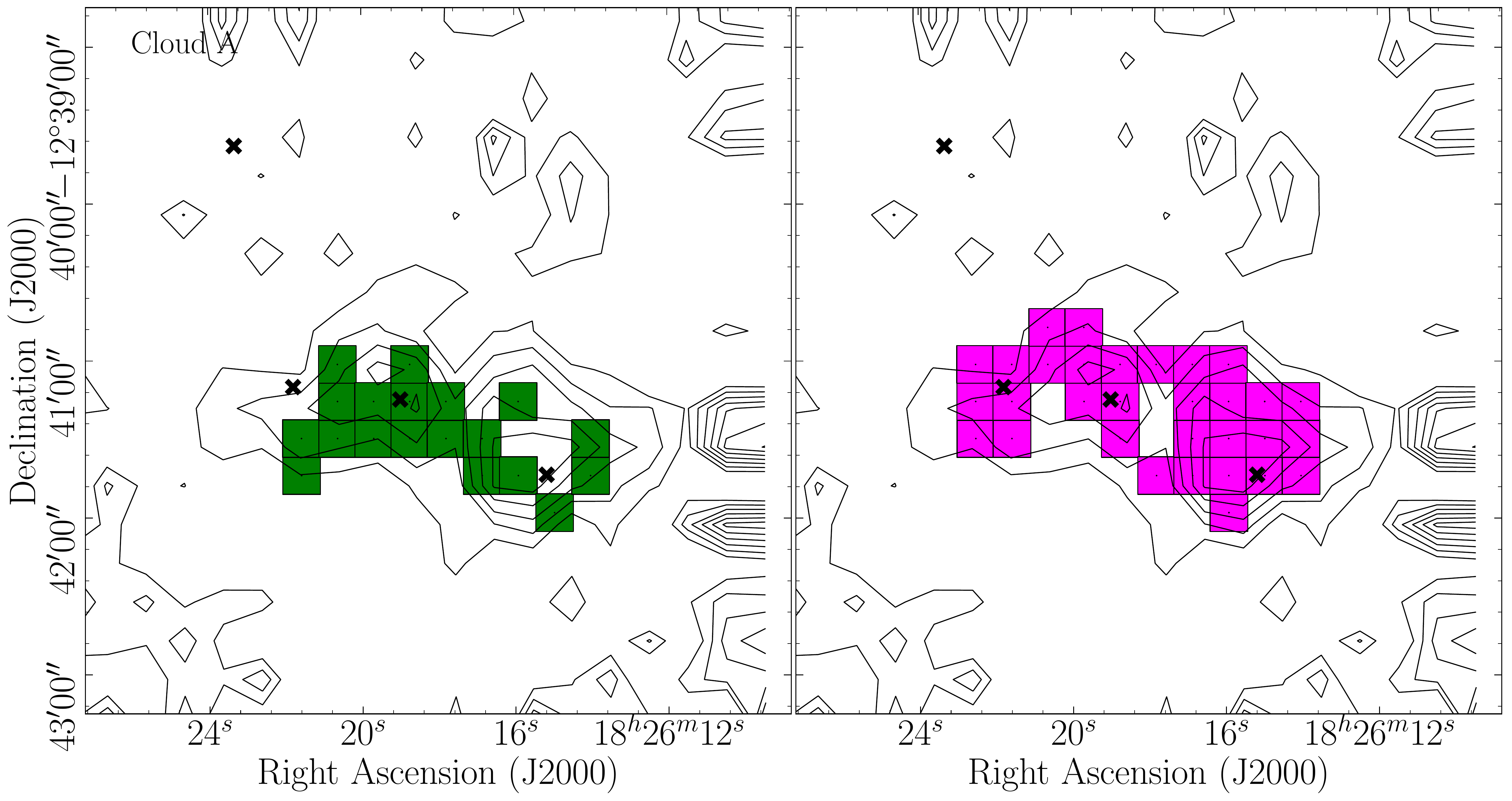}
    \includegraphics[scale=0.25,trim = 0.3cm 0cm 0cm 0cm, clip=True]{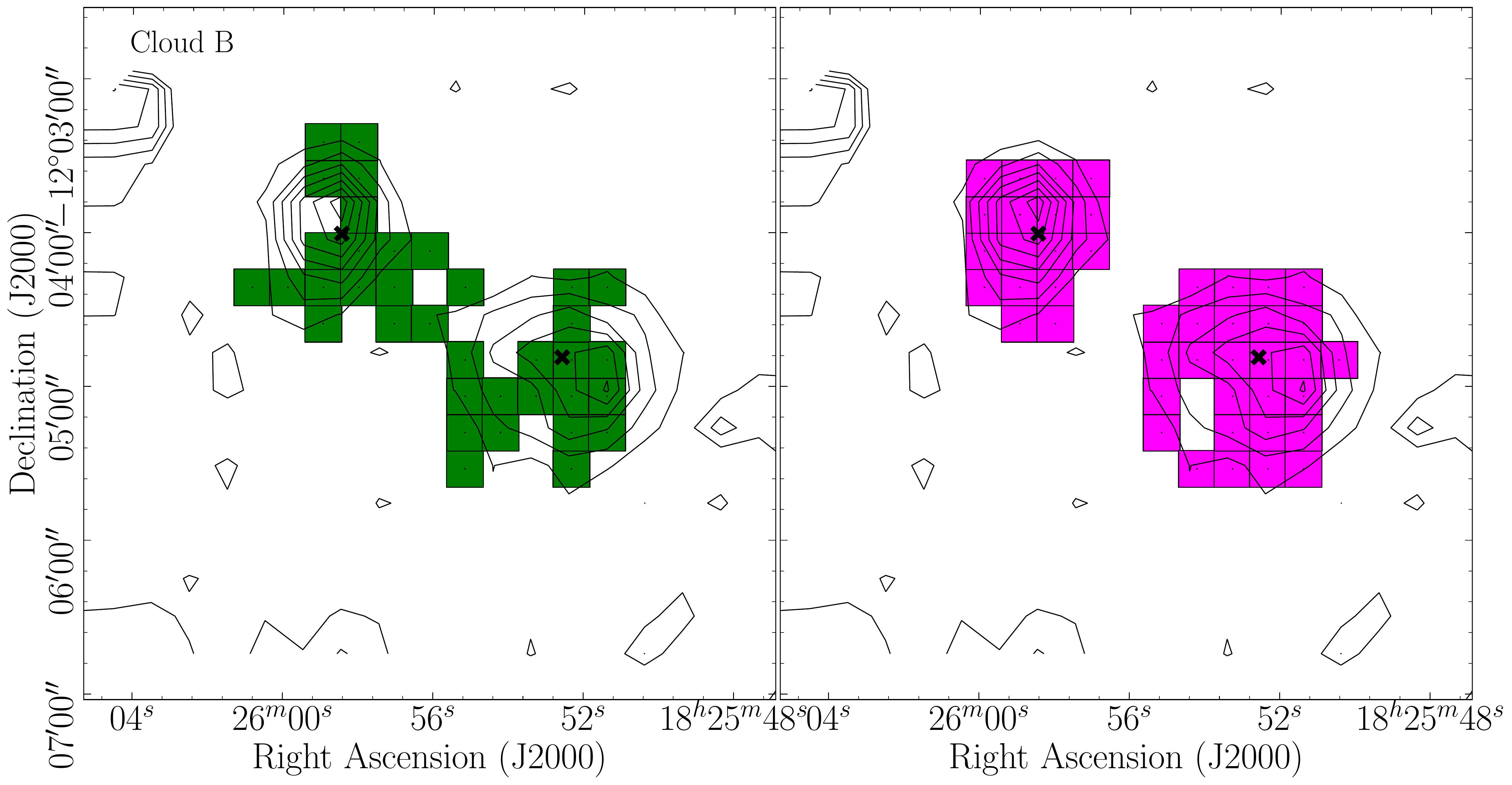}
    \includegraphics[scale=0.25,trim = 0.3cm 0cm 0cm 0cm, clip=True]{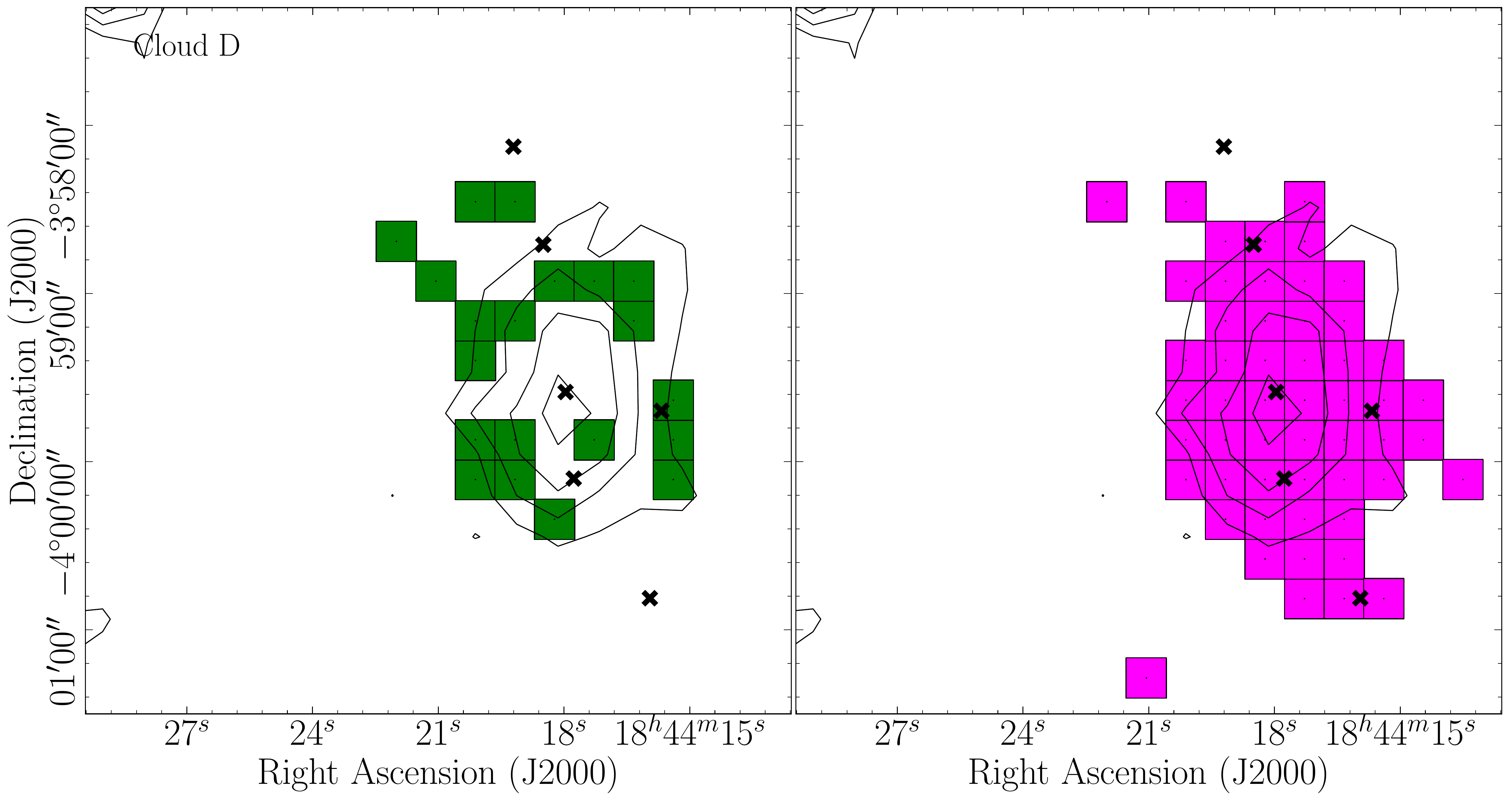}
    \caption{Spatial distributions of the narrow (green squares, left-hand panels) and broad (magenta squares, right-hand panels) SiO line width components toward cloud A, B and D from top to bottom respectively. Black contours corresponds to SiO integrated intensity maps as presented in Figure~\ref{siomaps}. Black crosses indicated the massive cores positions as in \citet{butler2012}.}
    \label{SiOwidthmap1}
\end{figure*}

\begin{figure*}
    \includegraphics[scale=0.25,trim = 0.3cm 0cm 0cm 0cm, clip=True]{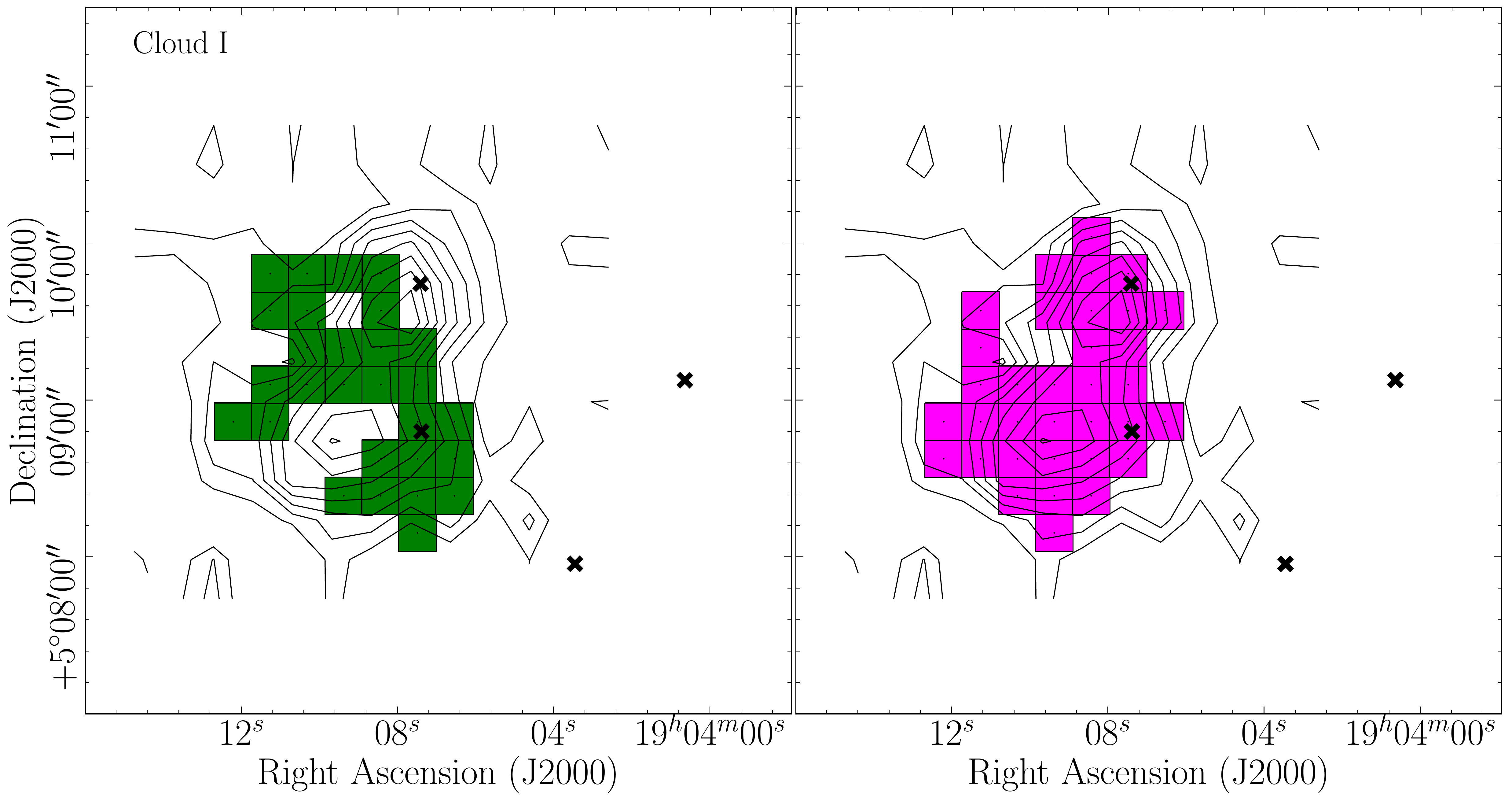}
    \includegraphics[scale=0.25,trim = 0.3cm 0.5cm 0cm 0cm, clip=True]{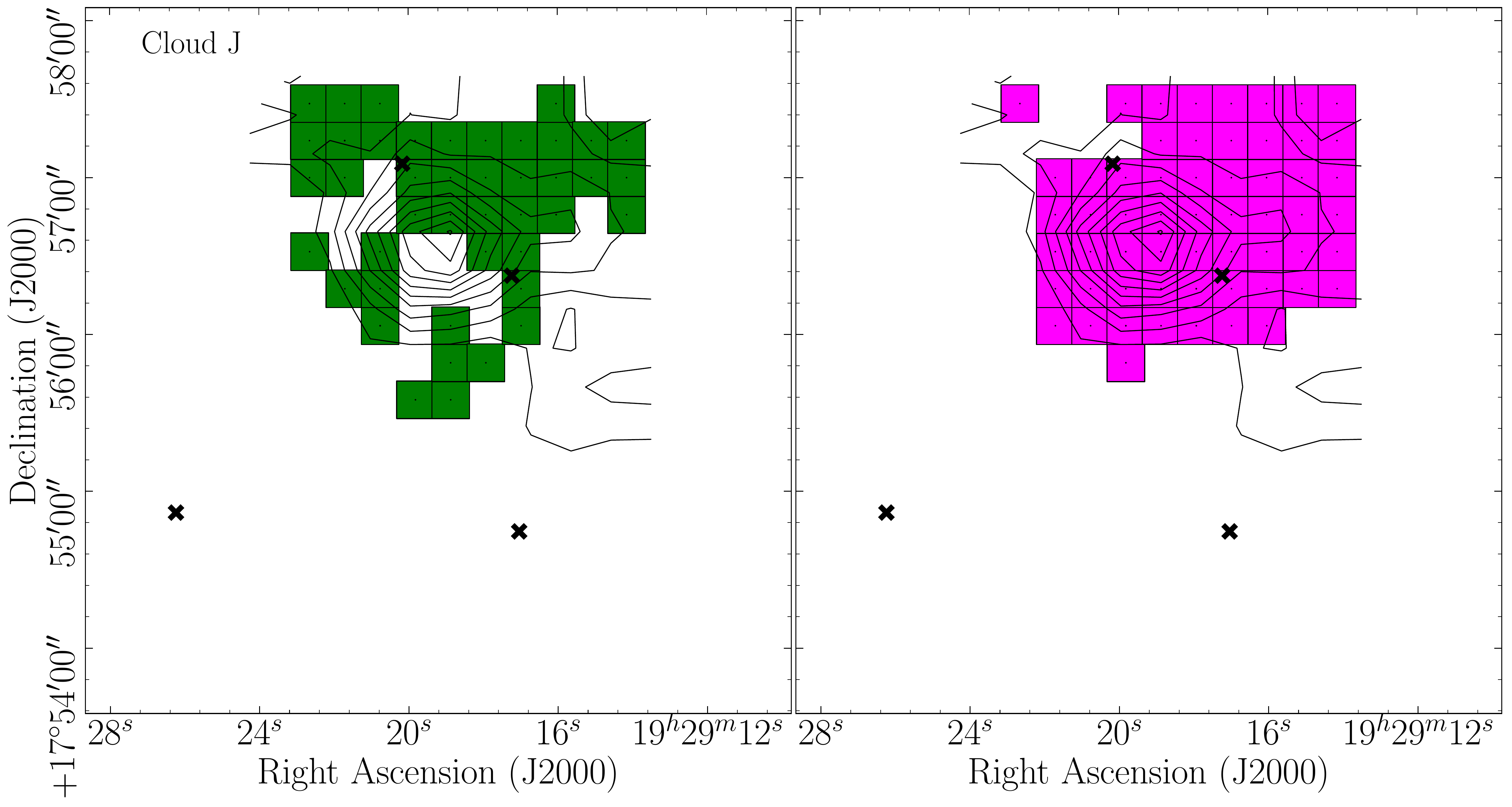}
    \caption{Spatial distributions of the narrow (green squares, left-hand panels) and broad (magenta squares, right-hand panels) SiO line width components toward cloud I (top panel) and J (bottom panel). Black contours corresponds to SiO integrated intensity maps as presented in Figure~\ref{siomaps}. Black crosses indicated the massive cores positions as in \citet{butler2012}.}
    \label{SiOwidthmap2}
\end{figure*}

In Figures~\ref{SiOwidthmap1} and ~\ref{SiOwidthmap2}, we show the spatial distribution of the broad (magenta squares in right-hand panels) and narrow (green squares in left-hand panels) SiO line width components toward clouds A, B, D, I and J and compare them with the global SiO morphology (black contours as in Figure~\ref{siomaps}) across the clouds. 

\noindent
Towards cloud A, the two line width components only coexist in correspondence of the massive core positions. The broad component extends beyond the massive cores (toward the northern region of the cloud), while the narrow component is mainly located below them (toward the southern region of the cloud). The narrow SiO component has mean intensity-weighted line widths of $\sim$2 \kms \space and, as seen in Figure~\ref{SiOspectra} (bottom panel in cloud A), positions of isolated narrow SiO emission can be identified.\\
\noindent
Toward cloud I, the narrow and broad emission components show mirrored distributions with respect to the core positions. The broad emission is mainly located around the two massive cores J4 and J5 and extends from north-west to south-east while the narrow emission lies in between the two cores and extends from north-east to south-west. Similarly to cloud A, regions of spatially isolated narrow emission can be identified across the cloud (Cloud I, bottom panel in Figure~\ref{SiOspectra}), with line widths always in the range 3 to 4 \kms.\\
\noindent
Toward clouds B, D and J, the two line width components are always coexistent and the narrow emission does not appear spatially isolated from the broad component. We note that the small percentage of narrow component found in cloud D, and to some extent also in cloud J, is distributed as a shell around the broad emission. \\

\noindent 
We note that we are confident that the very broad features detected in the line width distributions of the five IRDCs are real because all fitted components show peak intensities $>$3$\times$rms and integrated areas $>$3$\times$ the integrated rms, as imposed by {\sc SCOUSE}.

\subsubsection{The SiO velocity distributions}
\begin{figure}
\centering
\includegraphics[scale=0.8,trim = 1cm 1.5cm 10cm 3cm, clip=True]{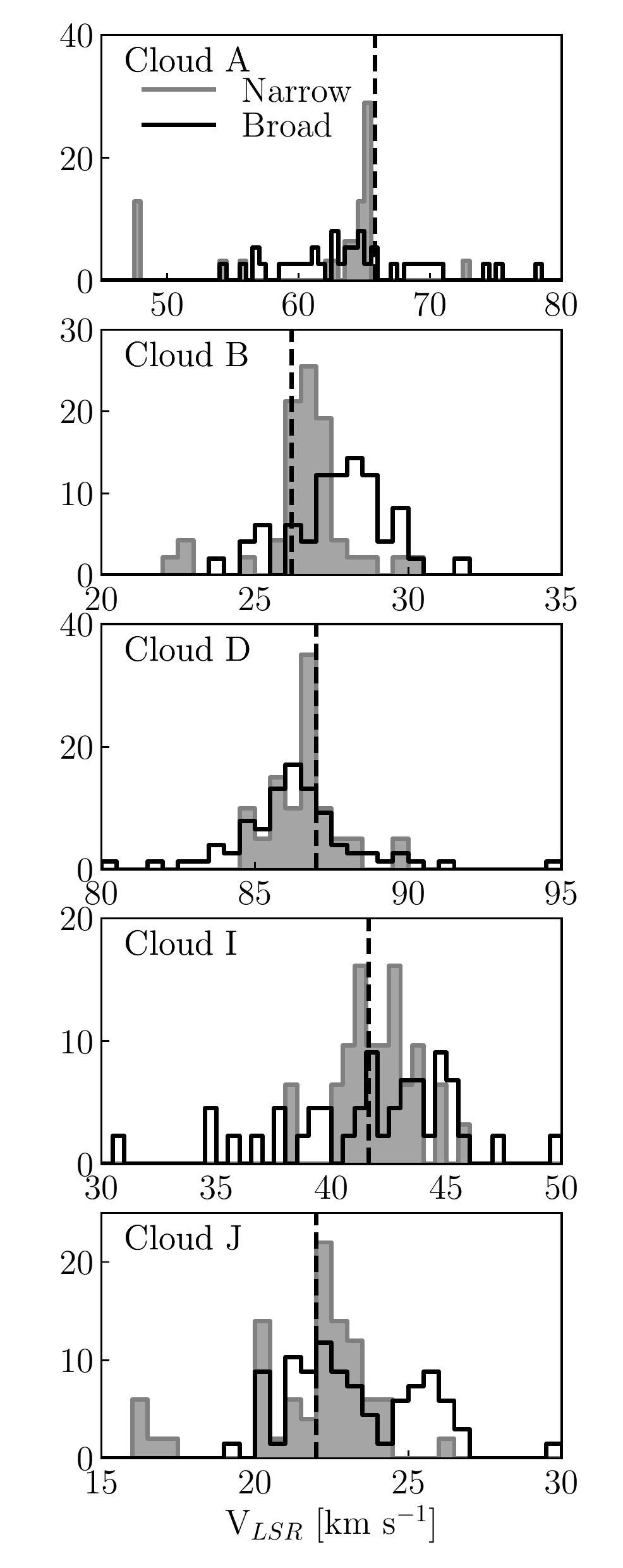}
\caption{Central velocity distributions of the SiO emissions obtained for cloud A, B, D, I and J and separately for the broad (empty black histograms) and narrow (grey filled histograms) emission components. The histograms show the percentage of lines having central velocity falling within each bin and normalised to the total number of narrow (broad) lines detected in each cloud. Bin size is 0.5 \kms \space corresponding to 1/3 of the mean intensity-weighted line width obtained for the dense gas tracers in \citet{cosentino2018}. The vertical dashed lines in all panels indicates the central velocity of the corresponding cloud \citep{simon2006b,rathborne2006}}
\label{SiOvelo}
\end{figure}

We complete the study of the SiO line profile across the five clouds by investigating the SiO centroid velocity distributions for the narrow and broad line width components, separately. Figure~\ref{SiOvelo} shows the centroid velocity distributions obtained for the narrow (grey filled histograms) and broad (black empty histograms) SiO emission for cloud A, B, D, I and J. For all the histograms, the bin size in the x-axis is 0.5 \kms \space (see Sec.~\ref{method}). In the y-axis we plot the percentage of emission lines having centroid velocity falling within the bin and normalised to the total number of narrow (broad) emission lines. The two velocity distributions have been superimposed to directly compare their kinematics. In all panels, the central velocity of the corresponding cloud, obtained from $^{13}$CO observations, is indicated \citep[vertical dashed lines;][]{rathborne2006}.

\noindent
The velocity distributions obtained for clouds B and J show a narrow line width component seen as a bright structure at the central velocity of the corresponding cloud, and a blue and/or red-shifted broad emission component. Toward cloud B, the broad emission is shifted by $\sim$3 \kms \space with respect to the central velocity of the cloud. Toward cloud J, the broad emission shows two defined velocity structures, one following the velocity distribution of the narrow emission and one red-shifted by $\sim$4 \kms. Toward cloud D, both the narrow and broad emission components show a symmetric spread in velocity of $\sim$20 \kms. Toward clouds A and I, the narrow emission is seen as a well defined structure with velocity dispersion of 2 to 3 \kms. The broad emission, however, does not show prominent structures and is spread across a velocity range of $\sim$10 \kms. Toward cloud I, the broad emission seems to be slightly concentrated at red-shifted velocities although not so prominently as observed toward clouds B, D and J. Finally, the narrow emission observed toward cloud I shows two emission peaks in the velocity distribution i.e. a first peak mainly associated with the central velocity of the cloud and a second peak, red-shifted by 1 to 2 \kms.\\\\

\noindent
In Figure~\ref{SiOspectra}, we have shown SiO spectra extracted across clouds A, B , D, I and J toward several positions, associated with both active and quiescent regions. Toward cloud A, we have selected positions A1, A2 and [$\alpha$(J2000) = 18$^h$26$^m$17.7$^s$ $\delta$(J2000) = $-$12$^{\circ}$41$^{\prime}$31.3$^{\prime\prime}$]. Toward this latter position, an isolated narrow component is clearly identified. Toward cloud B, broad SiO emission is seen toward the two core positions, B1 and B2, while no significant emission is detected toward the more quiescent region [$\alpha$(J2000) = 18$^h$25$^m$57$^s$ $\delta$(J2000) = $-$12$^{\circ}$04$^{\prime}$41$^{\prime\prime}$]. Toward cloud I, we have selected the positions of the massive cores I1 and I2 and the position [$\alpha$(J2000) = 19$^h$04$^m$10.6$^s$ $\delta$(J2000) = 5$^{\circ}$09$^{\prime}$15$^{\prime\prime}$]. Toward the latter position, the narrow SiO emission is isolated from the broad component, similarly to what is observed toward cloud A. However, it shows mean intensity-weighted line width of $\sim$3 \kms, slightly larger than those measured in cloud A and almost a factor of two broader than those observed in cloud G \citep[$\sim$1.6 \kms;][]{cosentino2018}. Toward clouds D and J, due to the compact structure of the emission we have selected a single position for each cloud corresponding to the SiO emission peak i.e. [$\alpha$(J2000) = 18$^h$44$^m$17.5$^s$ $\delta$(J2000) = $-$3$^{\circ}$59$^{\prime}$36$^{\prime\prime}$] toward cloud D and [$\alpha$(J2000) = 19$^h$29$^m$19$^s$ $\delta$(J2000) = 17$^{\circ}$56$^{\prime}$36$^{\prime\prime}$] toward cloud J. Both positions show very broad line profiles with line widths $\sim$10\kms.

\subsection{SiO Column Density}\label{abundances}

\noindent
By considering the spectra extracted across the clouds and shown in Figure~\ref{SiOspectra}, we use the software \textsc{MADCUBA} \citep{rivilla2016,martin2019} to estimate the SiO total column density values for the narrow and broad components toward the selected positions. 
We assume excitation temperature T$_{ex}$ of 10 K for the narrow component, consistent with values estimated toward cloud H from multi-line SiO analysis for the narrow component of SiO \citep{jimenezserra2010} and from other tracers in several works \citep{henshaw2014,jimenezserra2014} and T$_{ex}$ of 50 K for the broad component as estimated for shocked gas in molecular outflows \citep{jimenezserra2005}. We also note that the excitation temperature value assumed for the narrow emission component is consistent with those obtained in \citet{cosentino2018} from the narrow CH$_3$OH emission detected toward cloud G. Note that CH$_3$OH is also a good tracer of gas recently processed by shocks \citep{jimenezserra2005}.\\

\noindent
In Table~\ref{SiOabundance}, we report the N(SiO) for the broad and narrow emission, along with their ratios, toward all the selected positions across the clouds. For the narrow emission, we find N(SiO) in the range 3$\times$10$^{11}$ to 1.1$\times$10$^{12}$ cm$^{-2}$ for clouds A and B; from 5.7$\times$10$^{12}$ cm$^{-2}$ to 8.7$\times$10$^{12}$ cm$^{-2}$ in cloud D and J; 1 to 2$\times$10$^{12}$ cm$^{-2}$ in cloud I. For the broad component we find N(SiO) $\sim$5$\times$10$^{12}$ cm$^{-2}$ toward the active regions of cloud A and we measure an upper limit of $<$2$\times$10$^{11}$ cm$^{-2}$ toward the more quiescent region. Toward the massive cores in clouds B, D and J, the broad emission component shows N(SiO) in the range 1.4$\times$10$^{13}$-1.3$\times$10$^{14}$ cm$^{-2}$. Toward cloud I, the broad emission presents N(SiO) in the range 5-8$\times$10$^{12}$ cm$^{-2}$ toward the massive cores I1 and I2 and an upper limit of $<$2$\times$10$^{11}$ cm$^{-2}$ toward the more quiescent region. Finally, considering the estimated rms in the spectra of 9 mK (see Table~\ref{tab1}) in cloud E and line widths of 2 \kms \space and 10 \kms, we estimate upper limits of N(SiO)$\leq$ 4$\times$10$^{10}$ cm$^{-2}$ and N(SiO)$\leq$10$^{11}$ cm$^{-2}$ for the narrow and broad emission respectively. The ratio of the total SiO column densities between the narrow and broad components is $\leq$0.4 toward active regions. Toward the quiescent regions of clouds A and I, the SiO total column density of the narrow emission component is enhanced by a factor $\sim$2 to 5 with respect to that of the broad emission.\\

\noindent
The uncertainty on the SiO total column density values provided by \textsc{MADCUBA} is $\sim$10\%, as inferred by the  \textsc{SLIM} fitting tool within \textsc{MADCUBA}. In addition, the assumptions made for our analysis may introduce additional sources of uncertainty in the N$_{tot}$ estimates. For instance, already a factor of 2 increase in T$_{ex}$ will increase N(SiO) by a factor of $\sim$2. This is comparable to the enhancement between the narrow and broad component observed toward the quiescent regions of clouds A and I and above those observed in the other clouds. Hence, although our analysis suggests an enhancement of the narrow component toward more quiescent regions of the clouds, it is difficult to infer, from a single-transition analysis, the significance of such an enhancement. Hence a multi-transition analysis would need to be performed to better trace the excitation state of the shocked gas and thus obtain the SiO abundance \citep[as done in ][]{jimenezserra2010}.

\begin{table*}
    \centering
        \caption{SiO total column densities, and their ratios, measured for the narrow and broad components in several positions towards the six clouds.}
    \begin{tabular}{ccccc}
    \hline
    \hline
         Cloud & Position &\multicolumn{2}{c}{N(SiO)$\times$10$^{12}$ cm$^{-2}$} & N$_{narrow}$/N$_{broad}$\\
               &          & Narrow & Broad  &\\ 
    \hline           
         A    & A1   &1.0 &5.0 &0.2 \\
               & A2   &0.3 &5.0&0.06\\
               & 18$^h$26$^m$17.7$^s$ $-$12$^{\circ}$41$^{\prime}$31.3$^{\prime\prime}$ & 0.4 &$\leq$0.2 &$\geq$2\\
    \hline           
        B      & B1   &1.1 &14.0 &0.08\\
               & B2   &0.9 &130  &0.01\\
              & 18$^h$25$^m$57$^s$ $-$12$^{\circ}$04$^{\prime}$41$^{\prime\prime}$ &$\leq$0.007 &$\leq$0.2 & - \\
    \hline           
        D      & 18$^h$44$^m$17.5$^s$ $-$3$^{\circ}$59$^{\prime}$36$^{\prime\prime}$ &5.7 &18 & 0.32\\
    \hline    
        E      & $\cdots$ &$\leq$0.04 &$\leq$0.1 & -\\
    \hline    
        I      & I1  &1.7 &5.1 &0.33\\
               & I2  &2.0 &6.3 &0.32\\
               & 19$^h$04$^m$10.6$^s$ 5$^{\circ}$09$^{\prime}$15$^{\prime\prime}$ &1.0 &$\leq$0.2 &$\geq$5\\
    \hline           
        J      &19$^h$29$^m$19$^s$ 17$^{\circ}$56$^{\prime}$36$^{\prime\prime}$ &8.7 &27.0&0.32\\
    \hline    
    \end{tabular}
    \label{SiOabundance}
\end{table*}

\section{Discussion}\label{discussion}
The importance of cloud formation mechanisms and large-scale dynamics in the ignition of massive star formation in IRDCs can be tested, from an observational point of view, by analysing the kinematics and the spatial distribution of molecular emission across these sources \citep{jimenezserra2010,nguyen2013,jimenezserra2014,duartecabral2014,bisbas2018}. In cloud-cloud collisions the encounter of two pre-existing clouds and/or molecular filaments generates a shock that is predicted to be extended over a parsec-scales and to show velocities comparable to those observed in shear motions \citep[$\sim$10 \kms;][]{li2017}. The collision of such clouds or filaments can be induced either by the dynamics of the clouds orbiting the galactic plane \citep{tan2000,tasker2009,henshaw2013,inoue2013,jimenezserra2014,wu2015,colling2018} or triggered by external stellar feedback that sweeps up the interstellar medium material \citep{inutsuka2015,fukui2018,fukui2019,cosentino2019}. In both cases, signatures of such cloud-cloud collisions are imprinted in the kinematics of molecular tracers. Hence, by investigating the molecular gas content and its physical conditions, we can trace back the formation and processing history of the cloud. This is more challenging in sources with advanced levels of star formation activity because the cloud's pristine environment and hence the gas kinematics have already been affected.\\ 

\noindent 
In the following, we will discuss the likelihood for clouds A, B, D, I and J to have experienced a cloud-cloud collision event, based on the measured properties of the narrow and broad components observed for SiO, toward these clouds. We note that the order in which the clouds are discussed does not imply any evolutionary trend. Indeed, only for cloud H an age estimate has been given on the base of deuteration levels estimates across the cloud \citep[$\sim$3 Myrs;][]{barnes2018}. \citet{kong2017} investigated the presence of several deuterated species (especially N$_2$D$^+$) toward the massive cores within the clouds A, B, C, D, E, F and H. However, the N$_2$D$^+$ emission at a cloud spatial scale is currently not available and there is no evidence of a correspondence between the youth of a core and the youth of the hosting cloud. Hence, from such a study is very difficult to infer the relative evolutionary stage between the ten clouds in the sample.\\ By using the timescale for SiO depletion, we estimate a lower limit for the age of these clouds that is of the order of 10$^5$ years, i.e. the typical outflow lifetime. However, as mentioned above, the lifetime of these clouds are likely to be of the order of few Myrs when inferred from chemical clocks, such as the N$_2$D$^+$/N$_2$H$^+$ ratio. Due to the lack of any large scale N$_2$D$^+$ observations, it is currently not possible to establish the evolutionary stage of the clouds.\\ 
Among the clouds showing isolated narrow SiO emission, we speculate that cloud G \citep{cosentino2018,cosentino2019} is likely at the earliest evolutionary stage since no star formation activity is found toward this cloud. Following this argument, cloud A and I may be at an intermediate stage because they show both narrow and broad SiO emission, while clouds B, C, D, F and J are at a more evolved phase in their evolution due to the presence of only broad SiO emission and/or strong IR signatures of star formation activity toward the massive cores. However, this needs to be further investigated by studying the large scale emission from deuterated species.

\subsection{The SiO emission in Cloud B, D and J}
Toward the IRDCs B, D and J, the SiO narrow and broad emission components coexist everywhere. In all the three sources, the SiO emission shows compact morphology, spatially associated with massive cores previously identified within the clouds \citep{rathborne2006,butler2009,butler2012}. The broad components present blue and red-shifted structures in their velocity distribution (see Figure~\ref{SiOvelo}) and their kinematics show the same trend in their velocity distribution as for the narrow emission components. Most of the massive cores within the clouds show features of ongoing stellar activity \citep{chambers2009}. The cores B1 and B2 in cloud B and J1 in cloud J are associated with point sources at 24 $\mu$m and/or slightly extended emission at 4.5 $\mu$m (indicating the presence of H$_2$ shock-excited gas), indicating that star formation has already been ignited and that the sources are likely driving molecular outflows \citep{noriegacrespo2004,marston2004}. Furthermore, the cores show emission at 70$\mu$m, \footnote{https://irsa.ipac.caltech.edu/applications/Herschel/} further supporting the idea of deeply embedded protostars within the massive cores (Barnes et al. in prep). We also note that the typical spatial scales of the observed SiO emission toward clouds B and J are consistent with those expected even in massive molecular outflows \citep[i.e sub-parsec scales][]{beuther2002}.\\

\begin{figure}
    \centering
    \includegraphics[width=0.5\textwidth]{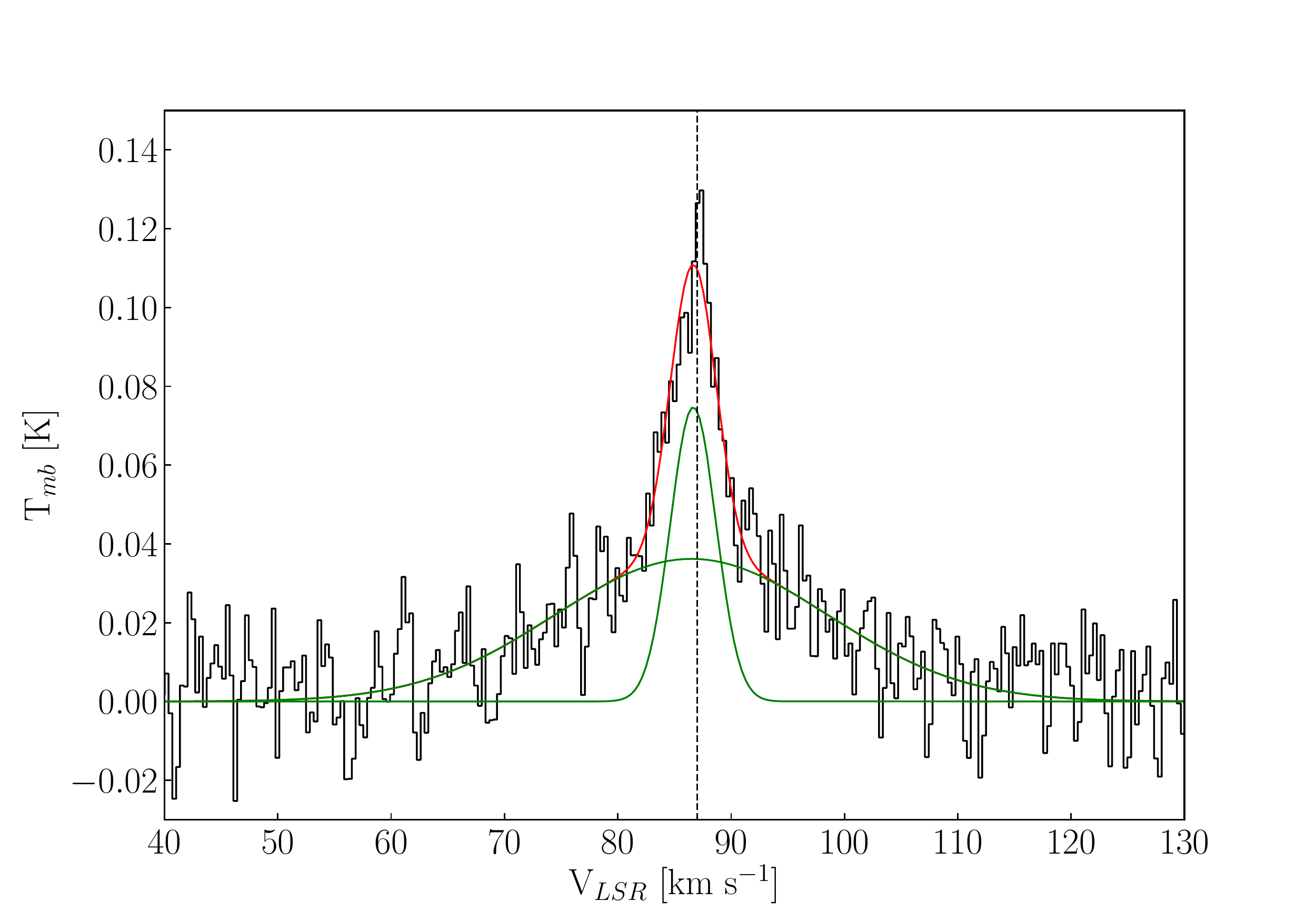}
    \caption{SiO line profile obtained toward cloud D by averaging the emission from the positions in which narrow SiO emission has been detected (green squares in Figure~\ref{SiOwidthmap1}). The multi-Gaussian fitting is indicated by the red line, while the single Gaussian components are indicated as green lines. The central velocity of the cloud is indicated as vertical dashed line.}
    \label{CloudDSpecNarrow}
\end{figure}

\noindent 
Toward cloud D, the massive cores are likely hosting deeply embedded protostars, as indicated by the presence of 70 $\mu$m and/or 24 $\mu$m, 8 $\mu$m point-like emission and "green fuzzy" emission at 4.5 $\mu$m \citep{chambers2009}. Hence, the SiO emission observed toward cloud D is likely to be associated with molecular outflows driven by protostars embedded in the active cores within the cloud.\\ 
\noindent
The SiO emission peak in cloud D coincides with the dust emission peak observed at 1.2 mm by \citet{rathborne2006} and the narrow emission component (see Figure~\ref{SiOwidthmap1}) is distributed as a shell around the blob-like morphology of the broad component. In Figure~\ref{CloudDSpecNarrow}, we show the SiO line profile obtained by averaging the emission in the positions where the narrow emission is detected (green squares in Figure~\ref{SiOwidthmap1}).

\noindent
We suggest that such a shell of narrow SiO emission may be arising from the post-shocked material decelerated by the interaction between the outflows and local dense clumps. This is similar to what was suggested by \citet{lefloch1998} as a possible origin for the narrow and bright SiO emission observed toward the molecular cloud NGC 1333. Assuming a typical molecular outflow lifetime of 10$^5$ years \citep{fukui1993} and terminal velocity of 10 \kms, and given the mechanical luminosity derived from the SiO emission of $\sim$7$\times$10$^{-3}$ L$\odot$ (L=2.6$\times$10$^{31}$ ergs s$^{-1}$), the mass required to decelerate the putative outflow toward cloud D is $\sim$0.08 M$\odot$. From Figure~\ref{siomaps}, the typical mass surface density at the cloud outskirts is 0.1 g cm$^{-2}$, corresponding to a mass of $\sim$2.2$\times$10$^2$ M$\odot$ when the same area of the outflow is considered. Hence, assuming that 100\% of the outflow kinetic energy is transferred to the cloud, the low density material at the cloud edges is enough to decelerate the outflow. We note that, it is likely that only a small fraction of the outflow kinetic energy \citep[$\leq$ 20\%;][]{arce2006} will be transferred to the cloud, further supporting the proposed scenario. The narrow SiO emission toward cloud D shows centroid velocity similar to that of the ambient gas, further supporting this scenario. We note that the assumed outflow lifetime is consistent with the SiO depletion time for typical IRDC density \citep[10$^4$ cm$^{-3}$;][]{martinpintado1992}.\\
Alternatively, SiO may be tracing the very first interaction between the MHD shocks associated with the putative molecular outflows and the surrounding clumpy material, similarly to what already proposed by \citet{jimenezserra2004} to explain the presence of narrow SiO ambient emission associated with the molecular outflows in L1448.

\noindent 
The angular resolution achieved in our observations (30\arcsec, corresponding to spatial scales of 0.3-0.9 pc at distances between 2 and 6 kpc) does not allow to spatially resolve multiple outflows from which the SiO emission toward cloud D may be arising. However, the cores B1, B2, D6 and D8 are part of the massive core sample investigated by Liu et al. in a forthcoming publication (Liu et al. in prep). The high-angular resolution images of the SiO(5-4) emission obtained by ALMA and investigated by these authors show the presence of compact and broad SiO(5-4) emission. This supports the idea that the cores are hosting embedded protostars driving molecular outflows.\\
We conclude that the SiO emission in clouds B, D and J is associated to ongoing star formation activity toward the three clouds, with the narrow and broad SiO emission tracing gas already processed by the MHD shock waves associated with molecular outflows \citep{martinpintado1992,codella1999,jimenezserra2005}. 

\subsection{The non-detection in Cloud E}
As presented in Section~\ref{results}, no SiO emission above the 3$\sigma$ ($\sigma$=0.1 K \kms) detection level is found toward cloud E. Consistently with our results, \citet{sanhueza2012} also report no SiO(2-1) emission toward the three massive clumps associated with the massive cores E1, E2 and E3.\\
\noindent 
The massive cores E1 and E2 do not show neither emission at 70 $\mu$m, 24 $\mu$m and 8 $\mu$m or green fuzzy emission at 4.5 $\mu$m \citep{chambers2009}. Toward these two positions across cloud E, Liu et al. in prep. report no SiO(5-4) emission associated with E2 and a 2$\sigma$ SiO(5-4) detection toward E1. The hint of emission detected toward E1 is extremely compact ($<$1\arcsec) and that may explain why it may have not been visible in our single-dish observations, due to a beam-dilution problem.\\

\noindent
The massive core E3 is spatially coincident with a point-like source seen at 70 $\mu$m, 24 $\mu$m, 8 $\mu$m and 4.5 $\mu$m emission \citep{chambers2009}, indicating that the source is likely hosting a deeply embedded protostar. The lack of SiO emission toward this core may be due either to beam dilution \citep[30\arcsec in this work and 38\arcsec in ][]{sanhueza2012} or to the fact that the source is in a stage evolved enough to not be driving molecular outflows.\\ 

\noindent 
Toward cloud E, no spatially widespread SiO emission is detected. This suggests two possible scenarios for the formation of the IRDC. The cloud may not be the result of a large scale shock interaction and hence alternative scenarios need to be considered as e.g., the gravitational collapse scenario \citep{heitsch2009,semadeni2019}. As a second possibility, the cloud may be the result of large scale shock interactions that may have occurred sometime in the past, so that their lifetime clearly exceeds the typical SiO freeze-out time but it is within the dynamical time scales required for the massive cores to become active. For the massive cores in cloud E, \citet{butler2012} report a volume-averaged H$_2$ number density n(H$_2$) $\sim$3$\times$10$^5$ cm$^{-3}$. For such a value of n(H$_2$), the SiO freeze-out time is estimated to be $\sim$10$^4$ years \citep{martinpintado1992}, while the free-fall time for the massive cores to collapse is estimated to be 3$\times$10$^5$ years. The results obtained toward cloud G in \citet{cosentino2019}, seem to support this scenario. Toward the IRDC G, indeed, we do not detect massive cores or evidence of ongoing stellar activity toward the region of the shock, suggesting that the formation of massive cores might be a consequence (and therefore subsequent) to the shock interaction.\\
\noindent 
Finally, a third scenario suggests that the putative large-scale shock interaction may be recent enough but the cloud may not be dense enough to probe the shock in SiO(2-1) (critical density 1.3$\times$10$^5$ cm$^{-3}$). However, the mass estimated by \citet{kainulainen2013} for cloud E ($\sim$2.9$\times$10$^4$ M$_{\odot}$) is a factor of 2 higher than the mass the authors estimated for cloud H ($\sim$1.7$\times$10$^4$ M$_{\odot}$) and almost a factor of 10 higher than that estimated for cloud G ($\sim$3$\times$10$^3$ M$_{\odot}$). Moreover, \citet{jimenezserra2010} estimated the SiO total column density toward cloud H to be in the range 5$\times$10$^{10}$ to 4$\times$10$^{11}$ cm$^{-3}$. This is comparable to the N(SiO) upper limit derived for cloud E and listed in Table~\ref{SiOabundance}. Hence, we suggest that due to the physical conditions of cloud E, any recent large-scale shock interaction should have been detected by its SiO emission.

\subsection{The case of cloud A and I}
The SiO emission detected toward clouds A and I shows a morphology and a kinematic structure different from the other sources of our sample. In fact, toward both clouds, the shock tracer morphology is extended across the whole cloud main filaments. This is opposite to what was found toward clouds B, D and J, where the SiO emission shows compact morphology around sites of ongoing star formation activity. In addition, although the SiO emission peaks are coincident with the position of the massive cores within clouds A and I, the narrow shock tracer emission is found to be partially detached both in morphology and kinematics from the broad component. This is opposite to what was observed toward the clouds of the sample with higher levels of star formation activity. In the following, we discuss possible mechanisms that could explain the observed broad and narrow SiO emission toward clouds A and I.

\subsubsection{The Origin of the SiO Emission toward Cloud A}
The massive cores A1 and A2 do not show evidence of associated point sources at 70 $\mu$m, 24 $\mu$m, 8 $\mu$m nor extended emission at 4.5 $\mu$m \citep{chambers2009}. This suggests that the cores are too young to be associated with IR signatures of star formation. However, the presence of broad SiO emission detected toward the core positions suggests that the core may be driving molecular outflows. This is supported by the fact that the mean intensity-weighted line widths and velocity distributions measured toward cloud A for the broad SiO emission component are comparable to those estimated for cloud B, D and J, where star formation is ongoing. In addition to the two massive cores, we also note the presence of bright compact features seen as a saturated hole in the mass surface density map of cloud A \citep{kainulainen2013}, toward the position $\alpha$(J2000) = 18$^h$26$^m$17$^s$ $\delta$(J2000)= $-$12$^{\circ}$41$^{\prime}$22$^{\prime\prime}$. This structure corresponds to a very bright point-like IR source, not explicitly reported by \citet{rathborne2006} but that may be associated with cloud A. Further investigations are needed to address the link between the cloud and such a source.\\
\noindent
All this points toward the idea that the SiO broad emission toward cloud A may be due to stellar feedback, probably driven by the massive cores A1 and/or A2 or the source at $\alpha$(J2000) =18$^h$26$^m$17$^s$  $\delta$(J2000) = $-$12$^{\circ}$41$^{\prime}$22$^{\prime\prime}$.\\ 

\noindent
As shown in Figures~\ref{siomaps}, the shock tracer emission toward cloud A is very widespread, covering a spatial scale of 4.2$\times$2.2 parsec$^2$, comparable to that observed toward cloud G in \citet{cosentino2018} and cloud H in \citet{jimenezserra2010} and more extended than the typical spatial scales observed in massive molecular outflows \citep{beuther2002}. For comparison, the spatial extent of the largest compact structure toward cloud B is 0.3$\times$0.7 parsec$^2$, more than a factor of two smaller than the SiO emission toward cloud A. We note that cloud A (d$\sim$4.8 kpc) is located further than cloud B (d$\sim$2.4 kpc) and hence the smaller SiO extent observed in cloud B is not due to distance effects.\\

\begin{figure}
    \centering
    \includegraphics[scale=0.3]{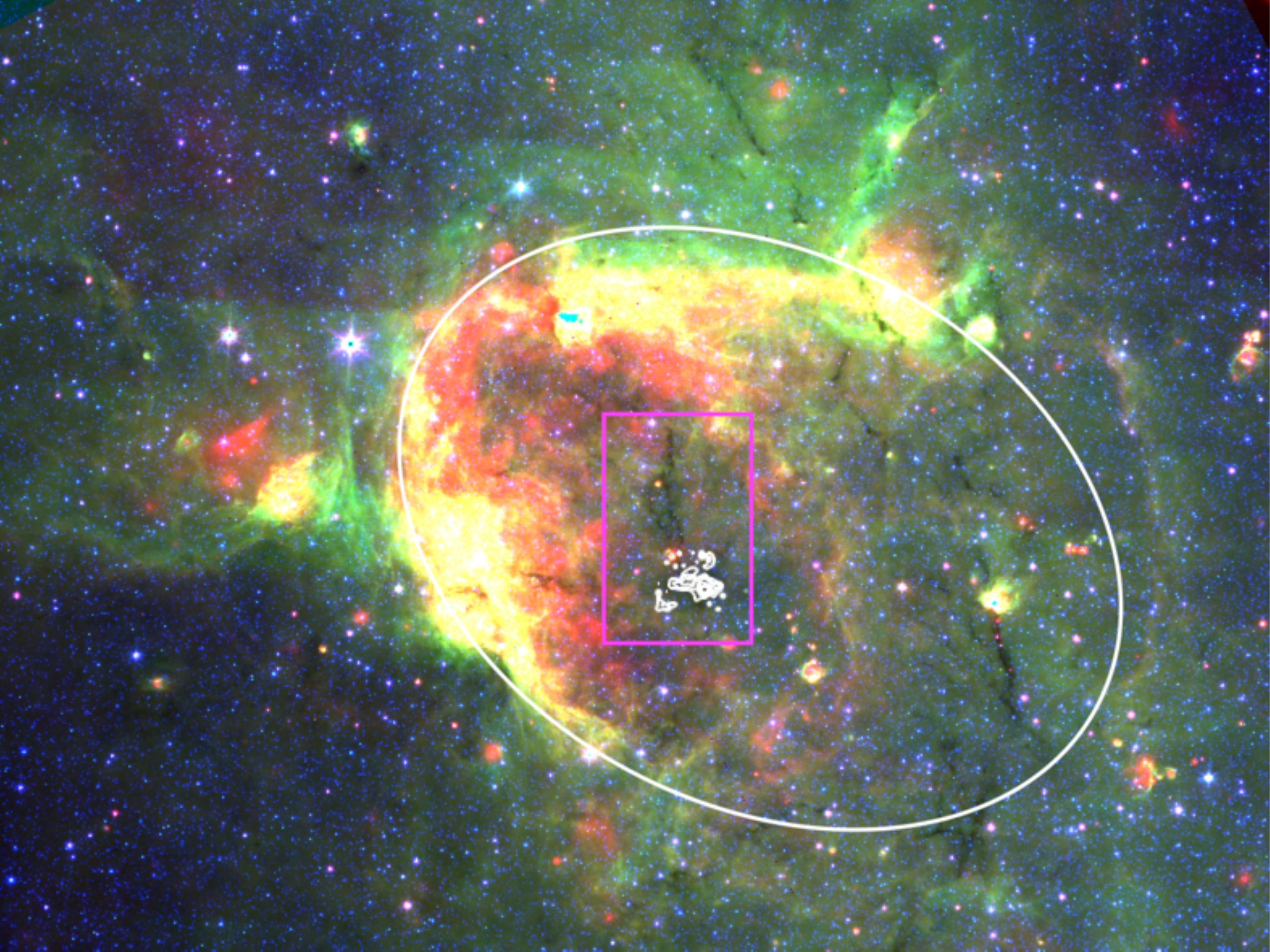}
    \caption{Three-color image of the Galactic Infrared Bubble N24 obtained from \textit{Spitzer} data. The 4.5 $\mu$m and 8$\mu$m emission have been obtained from the GLIMPSE Survey \citep{benjamin2003,churchwell2009} and are displayed in blue and green, respectively. The 24 $\mu$m emission is shown in red and has been obtained from the MIPSGAL Survey \citep{carey2009}. The white ellipse roughly indicates the position of N24, while the magenta rectangle highlights the position of cloud A. White contours show the SiO integrated emission levels as in Figure~\ref{siomaps}.} 
    \label{spitzercloudA}
\end{figure}

\noindent
Toward cloud A, we report the presence of narrow isolated SiO emission, extended in the southern region of the cloud (see Figure~\ref{SiOwidthmap1}) and with column densities higher by a factor of $\sim$2 with respect to that of the broad emission, in this region. Such a component is similar to the narrow and widespread SiO emission detected in cloud H by \citet{jimenezserra2010}. Furthermore, the mean intensity-weighted line widths of the SiO narrow emission are $\sim$2 \kms, similar to those measured toward cloud G \citep[$\sim$1.6 \kms;][]{cosentino2018} and toward cloud H \citep[$\sim$2 \kms;][]{jimenezserra2010}. Finally, as seen in Figure~\ref{SiOvelo}, the narrow SiO emission detected toward cloud A shows velocity distribution slightly blue-shifted ($\sim$1 \kms) with respect to the central velocity of the cloud. This behaviour is similar to that reported by \citet{cosentino2018} in cloud G.\\
\noindent
The enhanced narrow SiO emission, widespread across the southern part of the cloud, appears kinematically and spatially independent from the broad SiO emission and may be (to some extent) the result of a large-scale shock interaction. In Figure~\ref{spitzercloudA}, we inspect the cloud environment at large spatial scales by using \textit{Spitzer} images at multiple wavelengths. From Figure~\ref{spitzercloudA}, cloud A (magenta square) is encompassed by an arch-like structure identified as the Galactic Infrared Bubble N24 \citep[white ellipse;][]{churchwell2006,deharveng2010,simpson2012,kerton2013,li2019}.

\noindent
The bubble kinematic distance \citep[4.5 $\pm$ 0.2 kpc;][]{kerton2013} and central velocity \citep[64.5 $\pm$ 0.5 \kms;][]{kerton2013} are comparable to those inferred by \citet{simon2006a} for cloud A and reported in Table~\ref{tab1}. Recently, \citet{li2019} carried out a detailed multi-wavelength analysis of N24 and identified several clumps across the bubble with evidence of active on-going star formation activity. The authors analyse the kinematic structure of the low-density gas tracer $^{13}$CO \citep[from the GRS Survey;][]{jackson2006} and found an extended clump of molecular material spatially coincident with the IRDC A and with central velocity compatible with that of both the bubble and the cloud \citep[e.g see Figure 3 in ][]{li2019}. In addition, the $^{13}$CO observations from the GRS survey (Henshaw et al. in prep.) also show the presence of a molecular gas flow approaching the cloud across the same region where narrow isolated SiO emission is found toward cloud A.\\
\noindent
The lack of extended 8 $\mu$m emission across the IRDC \citep{li2019} suggests that the cloud is not being heated by the stellar wind and/or strong ultraviolet radiation field associated with the bubble and conclude that the cloud is likely located either in front or behind the bubble. If this is the case, cloud A may be interacting with the shock front layer of the expanding bubble or with the flow of molecular gas observed in $^{13}$CO, swept by the expansion of the H{\small II} region into the ISM. Such a scenario is similar to what has been reported toward cloud G \citep{cosentino2018,cosentino2019}.\\
Finally, no evidence of point like 8 $\mu$m sources is found in the region of cloud A where the isolated narrow SiO emission is detected. This indicates that the narrow SiO emission is likely not due to molecular outflows driven by deeply embedded sources. Hence, we support the idea that the collision between the molecular gas associated with the cloud and this additional flow associated with the H{\small II} region is (at least partially) responsible for the observed narrow SiO emission toward cloud A.
We shall further investigate this with interferometric observations and a detailed analysis of the IR emission at multiple wavelengths toward the cloud.

\subsubsection{The SiO Emission toward Cloud I}
\begin{figure}
    \centering
    \includegraphics[scale=0.28,angle=-90]{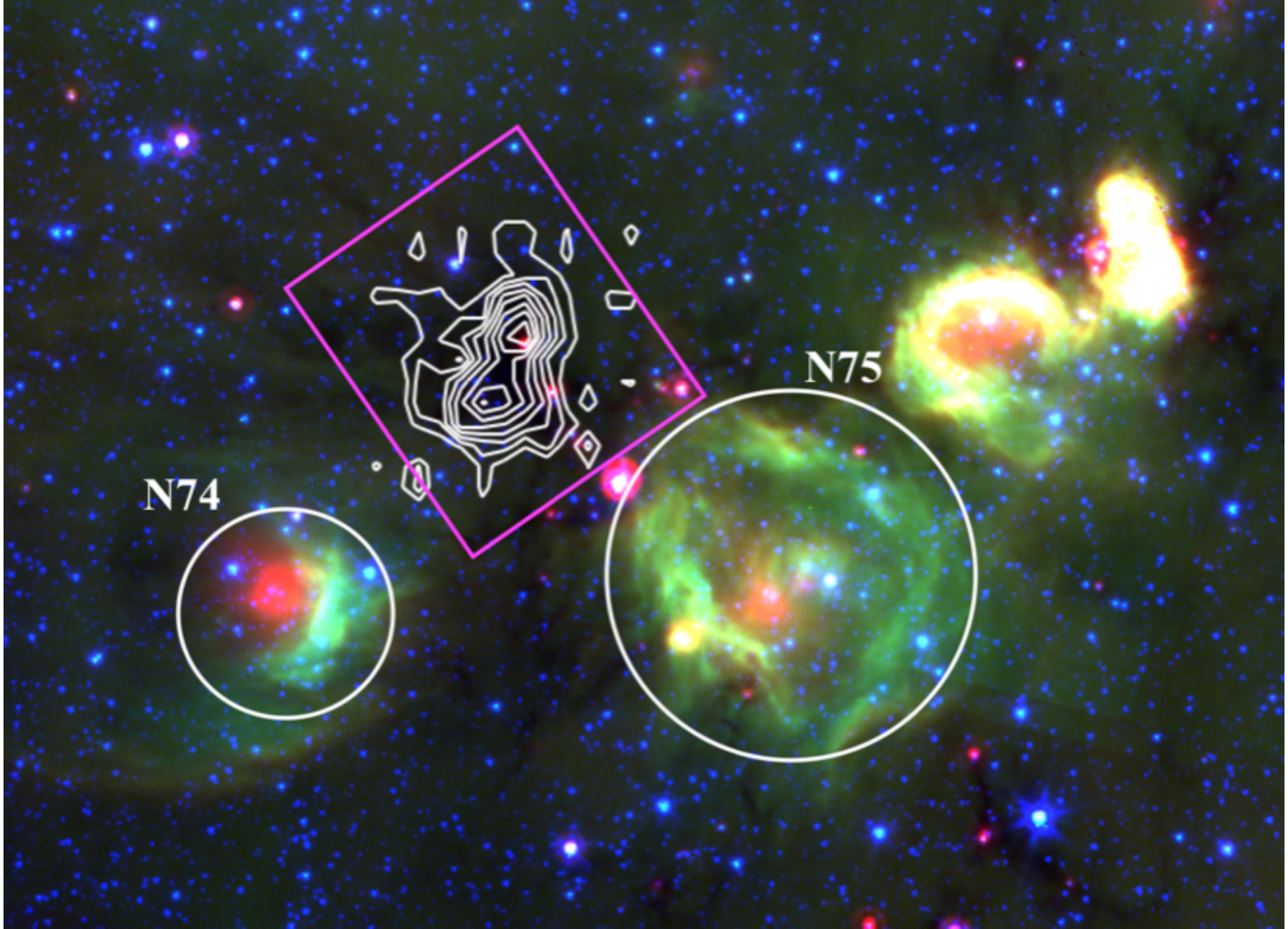}
    \caption{Three-color image of cloud I as surrounded by the two H{\small II} regions N74 and N75 (white circles). The 4.5 $\mu$m and 8$\mu$m emission have been obtained from the GLIMPSE Survey \citep{benjamin2003,churchwell2009} and are displayed in blue and green respectively. The 24 $\mu$m emission is shown in red and has been obtained from the MIPSGAL Survey \citep{carey2009}. The magenta rectangle indicates the position of cloud I. White contours show the SiO integrated emission levels as in Figure~\ref{siomaps}} 
    \label{spitzercloudI}
\end{figure}

Toward the IRDC I, the core I1/MM1 is known to be driving a molecular outflow and it hosts evidence of infall motion \citep{lopezSepulcre2010}. In contrast, the core I2 has been classified as quiescent by \citet{chambers2009}. From this, it is not surprising that the bulk of the broad SiO emission component is found toward the south of cloud I, where I1 is located (see Figure~\ref{siomaps}). Hence, the broad SiO emission detected toward the IRDC I is likely associated with star formation feedback driven by the core I1.\\
\noindent
The narrow SiO emission component detected toward cloud I shows mean intensity-weighted line widths of $\sim$3 \kms, higher that those observed in cloud G, H and A and comparable to those reported toward clouds C, F and J. The line width distributions of the narrow SiO emission does not show very bright peaks and appear to be similar to those observed toward clouds D and J (see Figure~\ref{SiOwidth}). However, isolated narrow SiO emission is detected toward regions of the cloud located in between the two massive cores. In addition, as seen in Figure~\ref{siomaps}, the SiO emission toward cloud I is widespread across an area of 1.3$\times$1.7 parsec$^2$, more extended than the typical SiO emission observed in molecular outflows. Finally, the column density values measured for the narrow SiO emission toward quiescent regions are enhanced by a factor of $\sim$5 with respect to the column densities measured for the broad component. All this seems to indicate that, although on-going star formation activity has already affected the cloud environment, an additional mechanism may be partially responsible for the widespread and enhanced narrow SiO emission observed toward the cloud. Similarly to what reported in the previous Section for cloud A, in Figure~\ref{spitzercloudI}, we investigate the large-scale environment of cloud I, by using \textit{Spitzer} images at multiple wavelengths.\\ 

\noindent
As shown in Figure~\ref{spitzercloudI}, Cloud I lies in between two known H{\small II} regions, G38.91-0.44 (or N74) and G39.30-1.04 (or N75). The three objects show similar central velocity of $\sim$40 \kms \space and are located at a similar kinematic distance of 2.9 kpc \citep{Du2008}. In \citet{Xu2013}, evidence of a possible interaction between the cloud and the expanding bubbles of the two H{\small II} regions is presented by means of multi-wavelengths observations. The authors suggest that the expansion of the two bubbles associated with the H{\small II} regions may have compressed the cloud and triggered the ignition of star formation activity toward I1. The spatial distribution of the narrow SiO emission shown in Figure~\ref{SiOwidthmap2} shows an arch-like structure that coincides with the intersection between the projection of the two bubbles associated with the H{\small II} regions and hence seems to support the scenario proposed by \citet{Xu2013}. Hence, we suggest that the interaction between the cloud and the two nearby H{\small II} regions may be (partially) driving the narrow SiO emission detected toward cloud I. As for cloud A, no evidence of point like sources in the 8 $\mu$m Spitzer images is found in correspondence of the narrow isolated SiO emission. Hence, narrow SiO is not produced by embedded protostars. We suggest that the narrow isolated SiO emission detected toward cloud I is tracing the shock interaction between molecular flows pushed by the nearby H{\small II} regions.
Future observations at higher angular resolution coupled with a detailed analysis of the low density gas kinematics will help to further address such a scenario.

\section{Are there different types of cloud-cloud collisions?}\label{types}  
In \citet{cosentino2018}, \citet{cosentino2019} and in this work, we have reported a detailed study on the kinematics and spatial distribution of SiO emission toward a sample of nine IRDCs. Within the sample \citep{butler2009,butler2012}, cloud G shows evidence of an ongoing collision between the cloud and a flow of molecular gas pushed toward the IRDC by the nearby SNR W44. The interaction is observed in the form of a time-dependent MHD CJ-type shock \citep{cosentino2019} and it is seen to be enhancing the gas density by a factor $\geq$10. Therefore the shock induces post-shock gas densities compatible with those required for massive star formation. In this work, we have reported the presence of widespread, narrow and isolated SiO emission toward the two IRDCs A and I that may be partially associated with the putative interaction between the clouds and nearby H{\small II} regions. In addition, \citet{jimenezserra2010} also report the presence of narrow and isolated SiO emission toward an additional source within the \citet{butler2009} sample i.e. the IRDC \irdcH, or cloud H. The study performed by \citet{jimenezserra2010} and later works support the idea that cloud H, that is not located in the proximity of Galactic bubbles or SNRs, is indeed the result of a cloud-cloud collision event \citep{henshaw2014,bisbas2018,barnes2018}.\\

\noindent 
Besides our sample, evidence of cloud-cloud collisions have been reported toward several sources by means of low density gas tracers \citep[e.g., $^{13}$CO emission; ][]{dewangan2018,kohno2018,tokuda2019,fujita2020}, shock  tracers \citep[SiO and CCS emission; ][]{nguyen2013,nakamura2015,louvet2016} and dense gas tracers emission \citep[e.g., H$^{13}$CO$^+$; ][]{dhabal2018}. Very recently, signatures of cloud-cloud collisions triggered by external stellar feedback have also been reported by \citet{dhanya2020} toward the S147/S153 complex.\\

\noindent
All these results suggest that, along with cloud-cloud collisions due to the natural motion of molecular clouds across the Galactic plane, i.e. \textit{natural} cloud-cloud collisions, filament collisions induced by stellar feedback may represent an efficient mechanism for triggering star formation in IRDCs. In these \textit{stellar feedback cloud-cloud collisions}, stellar feedback sweeps up the surrounding molecular material and pushes it toward pre-existing nearby molecular dense structures i.e., molecular clouds, dense clumps. The collision between the pushed material and the dense structure may initiate star formation. Indeed, recent simulations have shown that even the strong clumpy ejecta from SNs can penetrate to distances up to 1 parsec into molecular clouds \citep{Pan2012}.\\
The presence of SiO emission associated with collisions induced by mechanical stellar feedback depends on the nature of the stellar feedback itself. Strong events such as SNRs, H{\small II} regions and strong stellar winds, carry mechanical energies of the order of 10$^{49}$ ergs \citep{Tielens2005}. For typical molecular clouds of 10$^3$ M$_{\odot}$, the associated shock velocity is $>$30 \kms, enough to sputter dust grains and to inject SiO into the gas phase \citep{jimenezserra2008,nguyen2013}. We note that these shock velocities have been estimated by assuming that all the mechanical energy from the stellar feedback is transferred to the molecular clouds. It is likely that only part of this mechanical energy is transferred to the cloud. Moreover, in the case of SNRs and H{\small II} regions, the shock velocity maybe depends on the velocity of the expanding shell e.g., $\sim$ 10 \kms \space for H{\small II} regions \citep{Tielens2005}.\\ 
Stellar feedback cloud-cloud collisions show physical structure similar to those expected in natural cloud-cloud collisions but, as suggested by the extent of the shock interaction toward cloud G ($\sim$ 1 parsec), they may occur at smaller spatial scales i.e. parsec scale vs multiple parsec scale \citep{tan2000,tasker2009,wu2015,wu2016}. Hence, stellar feedback cloud-cloud collisions may be responsible for igniting star formation within the clouds and for helping to shape their filamentary structures, but it seems unlikely that they concur in the assembly of the IRDC itself and natural cloud-cloud collisions need to be further studied.\\

\noindent
\section{Conclusions}\label{conclusions}
In this work, we used single-dish IRAM 30m observations to analyse the spatial distribution, kinematic structure and line profiles of the SiO emission across the six IRDCs \irdcA, \irdcB, \irdcD, \irdcE, \irdcI \space and \irdcJ \space (clouds A, B, D, E, I and J respectively) and obtained the following results:

\begin{itemize}
    \item[i)] Of the six clouds, we only  detect significant SiO emission toward clouds A, B, D, I and J. In cloud E, the shock tracer emission is below the 3$\sigma$ detection level toward the whole extent of the area covered in our observations.\\ 
   
    \item[ii)] Toward clouds B, D and J, the SiO emission is spatially organised in blob-like structures whose positions are coincident with those of active massive cores previously identified within the clouds. On the contrary, the SiO emission toward clouds A and I shows a widespread morphology, extended over a parsec-scale and following the filamentary structure of the clouds as seen in extinction.\\
       
    \item[iii)] Across the five clouds, the SiO emission shows both a narrow ($\leq$ 5 \kms) and a broad line width components each accounting for $\sim$50\% of the total emission lines in clouds A, B, I and J and for nearly 80\% of broad emission lines in cloud D. Toward clouds B, D, and J the broad and narrow components are coexistent and are both spatially associated with the massive cores within the clouds. This indicates a common origin of the two line width components that are likely probing gas affected by the MHD shocks associated with ongoing star formation activity.\\ 
    
    \item[iv)] Toward clouds A and I, isolated narrow SiO emission is found toward the more quiescent regions across the IRDCs. The narrow and broad line width components show very different central velocity distributions and do not present prominent wing-like structures.\\ 
    
    \item[v)] Cloud A and cloud I are found to be spatially coincident and at similar kinematic distances of the Galactic bubble N24 and the H{\small II} regions N74 and N74, respectively. Due to the spatial morphology of the narrow emission component and to its kinematic structure, we speculate that the SiO emission toward clouds A and I may be tracing the ongoing interaction between the clouds and the flows of molecular gas pushed away by the expanding bubbles. This is supported by the velocity and spatial distributions of the narrow component.\\ 
    
    \item[vi)] Alternatively, the presence of low-mass star populations, undetected in \textit{Spitzer} images at multiple wavelengths, and associated with the clouds may be responsible for the narrow SiO emission across clouds A and I. Finally, projection effects may be responsible for the observed narrow SiO line profiles. 
\end{itemize}

\section*{Acknowledgements}
GC acknowledges support from a Chalmers Cosmic Origins postdoctoral fellowship. JCT acknowledges support from ERC project 788829 - MSTAR. ATB would like to acknowledge the funding provided from the European Union’s Horizon 2020 research and innovation programme (grant agreement No 726384). IJ-S has received partial support from the Spanish FEDER (project number ESP2017-86582-C4-1-R).
\section*{Data Availability}
The data underlying this article will be shared on reasonable request to the corresponding author.



\bibliographystyle{mnras}
\bibliography{IRDCs_gc} 





\bsp	
\label{lastpage}
\end{document}